\documentclass{aa}  
\usepackage{graphicx,natbib,psfig}  
\newcommand{\ltsima} {$\; \buildrel < \over \sim \;$}  
\newcommand{\gtsima} {$\; \buildrel > \over \sim \;$}  
\newcommand{\lta} {\lower.5ex\hbox{\ltsima}}  
\newcommand{\gta} {\lower.5ex\hbox{\gtsima}}

\defcitealias{capetti05}{Paper I}
\begin{document}  

\title{The host galaxy/AGN connection in nearby early-type galaxies
\thanks
{Based  on observations obtained at
the  Space  Telescope Science  Institute,  which  is  operated by  the
Association of  Universities for Research  in Astronomy, Incorporated,
under NASA contract NAS 5-26555.}.}
\subtitle{Is there a miniature radio-galaxy in every ``core'' galaxy?}
  
\titlerunning{A miniature radio-galaxy in every ``core'' galaxy?}  
\authorrunning{B. Balmaverde and A. Capetti}
  
\author{Barbara Balmaverde
\inst{1}
\and    
Alessandro Capetti \inst{2}}    
   
\offprints{B. Balmaverde}  
     
\institute{Universit\'a di Torino, Via Giuria 1, I-10125, Torino, Italy\\
\email{balmaverde@ph.unito.it}
\and
INAF - Osservatorio Astronomico di Torino, Strada
  Osservatorio 20, I-10025 Pino Torinese, Italy\\
\email{capetti@to.astro.it}}

\date{}  
   
\abstract{This is the second of a series of three papers exploring the
connection between  the multiwavelength 
properties of AGN in nearby early-type galaxies
and the characteristics of their hosts. 
We selected two samples with 5 GHz 
VLA radio flux measurements down to  1 mJy, reaching  levels of
radio  luminosity as  low as  10$^{36}$ erg s$^{-1}$. 
In \citetalias{capetti05}
we presented a study of the surface brightness profiles
for the 65 objects with available archival HST images 
out of the 116 radio-detected galaxies. We classified early-type galaxies into
``core'' and ``power-law'' galaxies, 
discriminating on the  basis of the slope of their  nuclear  brightness
profiles, following the Nukers scheme. 
Here we focus on the 29 core galaxies (hereafter CoreG).

We used HST and Chandra data to isolate their optical and X-ray
nuclear emission. The CoreG invariably host radio-loud
nuclei, with an average radio-loudness parameter of 
Log  R =  L$_{5\rm {GHz}}$ / L$_{\rm B}$ $\sim$ 3.6.
The optical and X-ray nuclear luminosities correlate with the
radio-core power, smoothly extending the analogous correlations 
already found for low luminosity radio-galaxies (LLRG) toward even lower
power, by a factor of $\sim 1000$, covering a combined range of 6
orders of magnitude. This supports the interpretation of a common
non-thermal origin of the nuclear emission also for CoreG.
The luminosities of the nuclear sources, 
most likely dominated by jet emission,
set firm upper limits, as low as L/L$_{\rm Edd} \sim 10^{-9}$ in both
the optical and X-ray band, on any emission from the accretion process.

The similarity of CoreG and LLRG when considering the distributions  
host galaxies luminosities and black hole masses,
as well as of the surface brightness profiles,
indicates that they are drawn from the same 
population of early-type galaxies. LLRG represent only the tip
of the iceberg associated with (relatively) high activity levels,
with CoreG forming the bulk of the population.

We do not find any relationship between radio-power and black hole
mass. A minimum black hole mass of $M_{BH} = 10^8 M_{\sun}$ is
apparently associated with the radio-loud nuclei in both CoreG and LLRG, 
but this effect must be tested on a sample  of less luminous
galaxies, likely to host smaller black holes. 

In the unifying model for BL~Lacs and radio-galaxies, 
CoreG likely 
represent the counterparts of the large population of low luminosity 
BL~Lac now emerging from the surveys at low radio flux
limits. This suggests the presence of relativistic jets 
also in these quasi-quiescent early-type ``core'' galaxies. 

} \maketitle

\section{Introduction.}
\label{intro}

The recent developments in our understanding of the nuclear regions of
nearby galaxies  provide us with a  new framework in  which to explore
the classical issue of the connection between host galaxies and AGN.

All evidence now points to the idea that most galaxies
host a supermassive black hole (SMBH) in their centers
\citep[e.g.][]{kormendy95} and that its mass is closely linked to the
host galaxies properties, such as the stellar velocity dispersion
\citep{ferrarese00,gebhardt00}.  This is clearly indicative of a
coevolution of the galaxy/SMBH system and it also provides us with
indirect, but robust, SMBH mass estimates for large sample of objects.
Furthermore, the innermost structure of nearby galaxies have been
revealed by HST imaging, showing the ubiquitous presence of singular
starlight distributions with surface brightness diverging as
$\Sigma(r)\sim r^{-\gamma}$ with $\gamma>0$ \citep[e.g.][]{lauer95}.
The distribution of cusp slopes \citep{faber97} is bimodal, with a
paucity of objects with $0.3<\gamma<0.5$.  Galaxies can then be
separated on the basis of their brightness profiles in the two classes
of ``core'' ($\gamma \leq 0.3$) and ``power-law'' ($\gamma \geq 0.5$)
galaxies, in close correspondence to the revision of the Hubble
sequence proposed by \citet{kormendy96}.

But despite these fundamental breakthroughs we still lack a clear
picture of the precise relationship between AGN and host galaxies.
For example, while spiral galaxies preferentially harbour radio-quiet
AGN, early-type galaxies host both radio-loud and radio-quiet
AGN. Similarly, radio-loud AGN are generally associated with the most
massive SMBH as there is a median shift between the radio-quiet and
radio-loud distribution, but both distributions are broad and overlap
considerably \citep[e.g.][]{dunlop03}.

In this framework, in two senses
early-type galaxies appear to be the critical class of
objects, where the transition between the two profiles classes occurs
(i.e. in which core and power-law galaxies coexist) and 
in which they can host  
either radio-loud and radio-quiet AGN. We thus started a
comprehensive study of a sample of early-type galaxies (see below for
the sample definition) to 
explore the connection between  the multiwavelength 
properties of AGN and the characteristics of their hosts. 
Since the 'Nuker' classification can only be obtained when the 
nuclear region, potentially associated with a shallow cusp,
can be well resolved, such a study must be limited to nearby galaxies.
The most compact cores will be barely resolved at a
distance of 40 Mpc (where 10 pc subtend 0\farcs05)
even in the HST images. Furthermore, high quality radio-images 
are required for an initial selection of AGN candidates. 

We then examined two samples of  nearby objects for
which  radio observations combining  relatively high  resolution, high
frequency and sensitivity  are available,  in order  to  minimize the
contribution from radio emission not related to the galaxy's nucleus
and confusion from background sources.
More specifically we focus on the samples of early-type galaxies 
studied by \citet{wrobel91b}  and \citet{sadler89} both observed with
the VLA at  5 GHz with a  flux limit  of
$\sim$  1 mJy.   The two  samples were  selected with  a  very similar
strategy. \citet{wrobel91a} extracted  a northern
sample  of  galaxies from  the  CfA  redshift survey  \citep{huchra83}
satisfying  the following  criteria: (1)  $\delta_{1950} \geq  0$, (2)
photometric magnitude  B $\leq$ 14; (3) heliocentric  velocity $\leq$ 3000
km s$^{-1}$, and (4) morphological  Hubble type T$\leq$-1, for a total
number of  216 galaxies.  \citet{sadler89} selected  a similar southern sample
of 116 E and S0 with  $-45 \leq \delta \leq -32$. 
The only difference between
the  two  samples  is  that  \citeauthor{sadler89} did  not  impose  a
distance limit.   Nonetheless,  the  threshold in  optical  magnitude
effectively limits the  sample to a recession velocity  of $\sim$ 6000
km s$^{-1}$. 

In \citet[ hereafter Paper I]{capetti05}, we focused on the 116 
galaxies detected in these VLA surveys to boost the fraction
of AGN with respect to a purely optically selected sample.
We used archival HST observations, available for 65 objects, to study their 
surface brightness profiles and to separate these early-type 
galaxies into core and power-law galaxies following the Nukers scheme, 
rather than on the traditional morphological classification (i.e. into E and
S0 galaxies). 
Here we focus on the sub-sample formed by the 29 ``core'' galaxies. 

We adopt a Hubble constant H$_{\rm o}=75$ km  s$^{-1}$ Mpc$^{-1}$.

\section{A critical analysis of the classification as core galaxies.}
\label{sersic}

In \citetalias{capetti05} we adopted the classification into power-law and
core galaxies following the scheme proposed by \citet{lauer95}.
We then separated early-type galaxies on the  basis of the 
slope of their  nuclear  brightness
profiles obtained using the Nukers law (i.e. a double power-law 
with innermost slope $\gamma$)
defining as core-galaxies all objects with $\gamma \leq 0.3$.
Since this strategy has been subsequently challenged by
\citet{graham03}, who introduced a different definition of
core-galaxies,
it is clearly important to assess whether the identification
of an object as a core galaxy is dependent on the fitting scheme adopted.

\citeauthor{graham03} argued that a S\'ersic model
\citep{sersic68}
provides a better
characterization of the brightness profiles of early-type galaxies.
In particular they pointed out that, among other issues, 
i) the values of the Nukers law
parameters depend on the radial region used for the
fit, ii) the Nukers fit is unable to reproduce the large scale
behaviour of early-type galaxies 
and, most importantly for our purposes, iii) the identification of
a core galaxy from a Nuker fit might not be recovered by a S\'ersic
fit. Conversely, they were able to fit power-law galaxies (in the
Nukers scheme), as well
as dwarf ellipticals \citep{graham03b}, with
a single S\'ersic law over the whole range of radii. 
They also suggested a new definition of 
core-galaxy as the class of objects 
showing a light deficit toward the center with respect to the S\'ersic
law \citep{trujillo04}.

\begin{figure*}
\centerline{
\psfig{figure=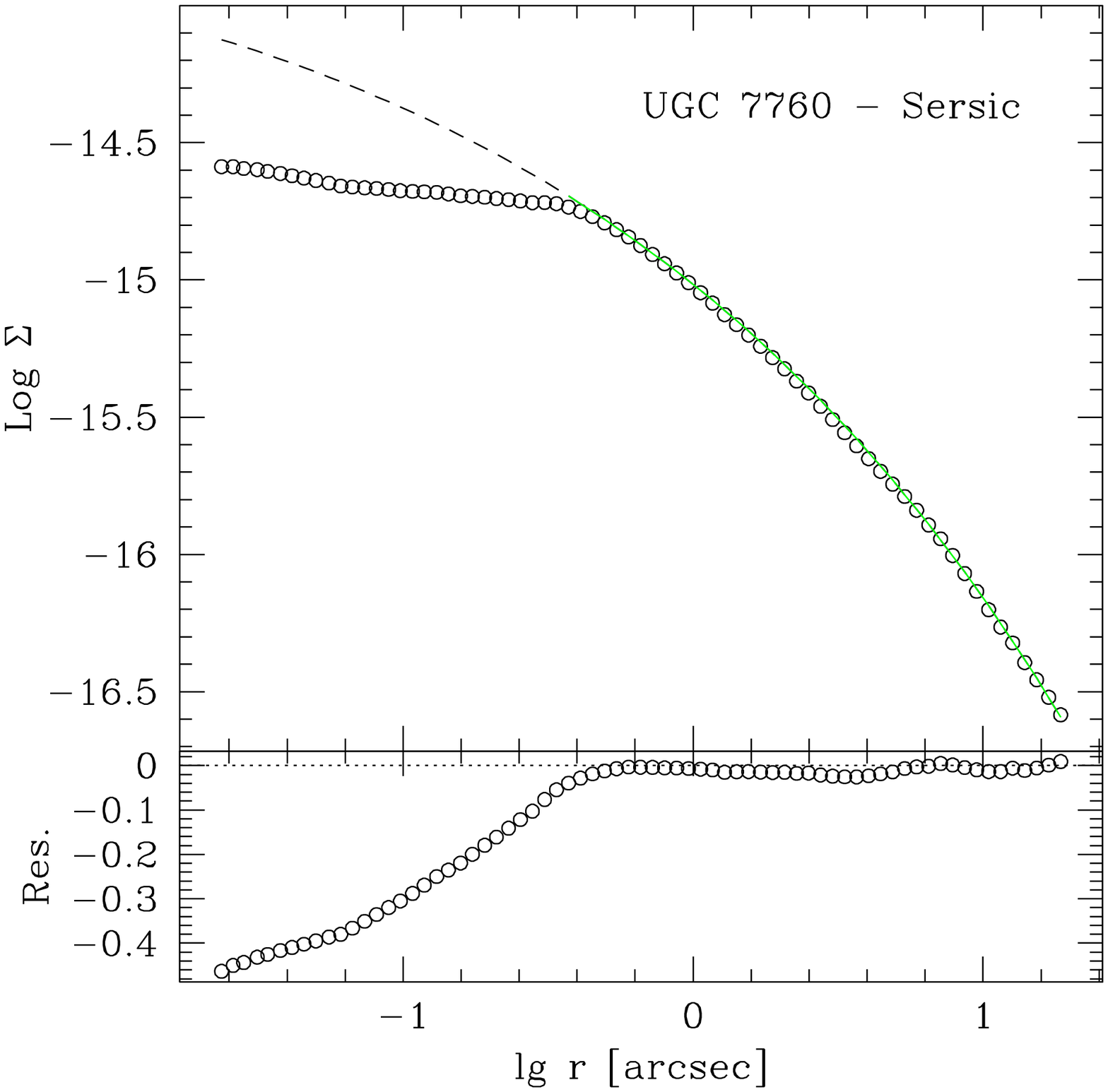,width=0.50\linewidth}
\psfig{figure=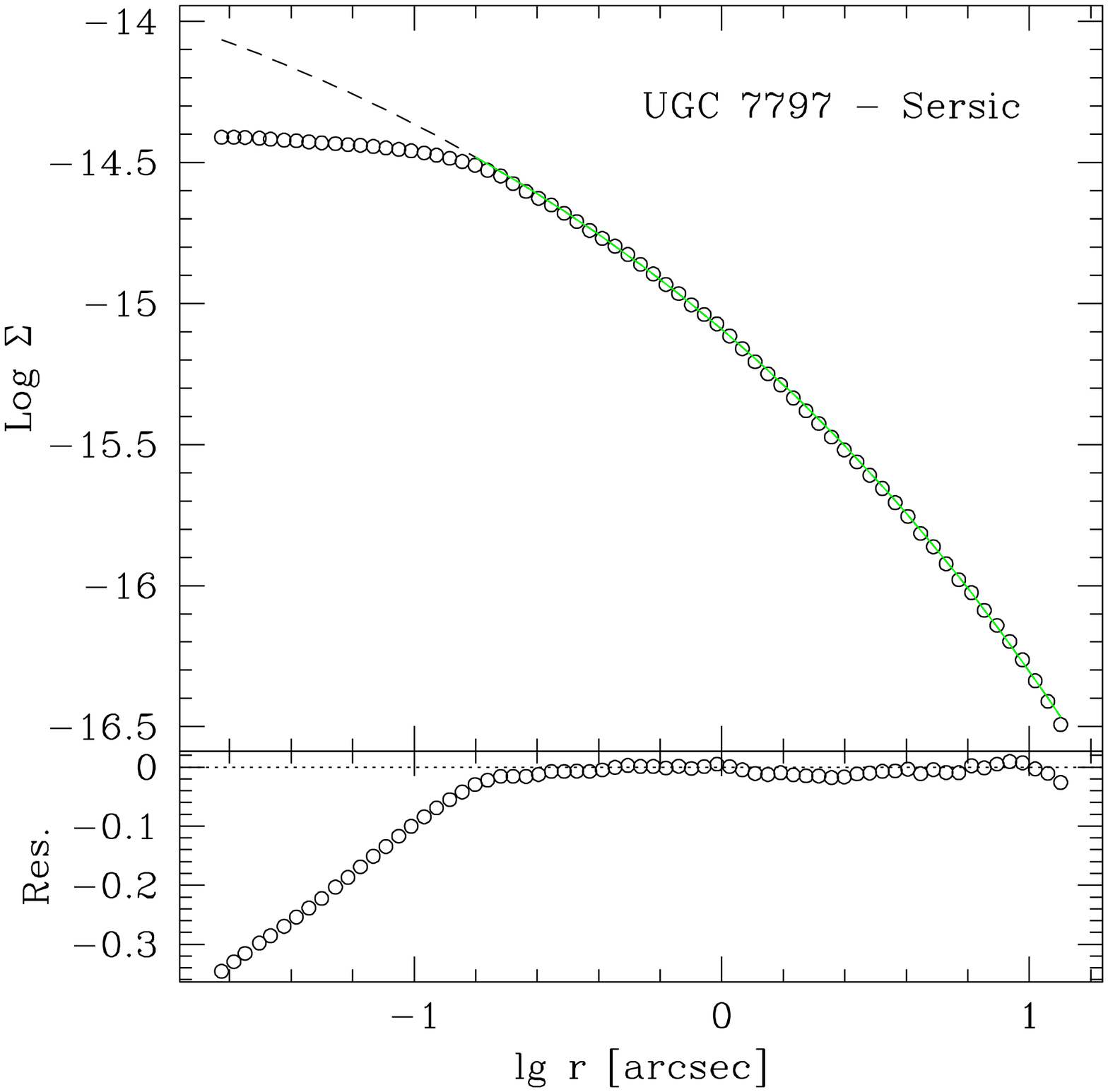,width=0.50\linewidth}
}
\caption{S\'ersic fits for the two core galaxies of our sample
with the smallest values for the core radius. A substantial central light 
deficit is clearly present in both objects, conforming to
the ``core'' classification in the Graham et al. scheme.}
\label{sersicfig}
\end{figure*}

In this context,
we discuss in detail  here the behaviour of the most critical
objects, i.e. the two core galaxies for which the Nuker
law returns the smallest values for the break radius,
namely UGC~7760 and UGC~7797 for which $r_b = 0\farcs49$
and $r_b = 0\farcs21$ respectively. We fit both objects with
a S\'ersic law. The final fits, shown in Fig. \ref{sersicfig},
were obtained iteratively, fitting the external regions
while flagging the innermost points
out to a radius at which the residual from the S\'ersic law exceeded
a threshold of 5 \%. The S\'ersic law  in general provides a remarkably good
fit to the outer regions, with typical residuals of $\sim$ 1\%, but 
a substantial central light deficit is clearly
present in both objects.
This indicates that both objects can be 
classified as core-galaxies in the Graham et al. scheme.

Using the brightness profiles for the core-galaxies for which
we obtained Nuker fits in \citetalias{capetti05} (14 additional objects) 
we obtained similar results. Very
satisfactory fits can be obtained with a S\'ersic law on the external
regions of these galaxies, but 
they all show an even clearer central light
deficit, as expected given the presence of well resolved shallow cores.

We conclude that, for the galaxies of our sample,
the objects classified as core-galaxies in the Nuker scheme 
are recovered as such with the Graham et al. 
definition.

\section{Basic data and nuclear luminosities}
\label{nuc}

\begin{table*}

\caption{
(1) optical name 
(2) Chandra observational identification number, (3) exposure time [ks],
(4) reference for the X-ray analysis (see below for the list),
(5) instrument and filter of the HST observation,
(6) optical nuclear flux [erg cm$^{-2}$ s$^{-1}$].}

\label{tabsample2}

\begin{tabular}{l | c c c | c r }

\hline		      	    
\hline		      	    
Name &\multicolumn{3}{|c|} {Chandra data summary} & \multicolumn{2}{c}{HST data summary}\\
     &  Obs. Id & Exp. time   & Ref. & Image & F$_{0}$   \\
\hline		      	    
UGC~0968 & --   &    --  &  --    & WFPC2/F814W    &   $<$4.0 E-14   \\
UGC~5902 & 1587 &  31.9  &  (1)   & WFPC2/F814W    &   $<$6.3 E-14   \\
UGC~6297 & 2073 &  39.0  &  (1)   & WFPC2/F814W    &   $<$5.0 E-14   \\
UGC~7203 & 3995 &  5.13  &  (1)   & WFPC2/F702W    &   4.0 E-14      \\
UGC~7360 &  834 &  35.2  &  (2)   & NICMOS/F160W   &   1.9 E-13      \\
UGC~7386 &  398 &  1.43  &  (3)   & NICMOS/F160W   &   1.6 E-13      \\
UGC~7494 &  803 &  28.85 &  (2)   & NICMOS/F160W   &   3.3 E-13      \\
UGC~7629 &  321 &  40.1  &  (4)   & WFPC2/F555W    &   1.4 E-14      \\
UGC~7654 & 1808 &  14.17 &  (2)   & NICMOS/F160W   &   5.1 E-12      \\
UGC~7760 & 2072 &  55.14 &  (5)   & WFPC2/F555W    &   6.8 E-14      \\
UGC~7797 &  --  &    --  &  --    & WFPC2/F702W    &   $<$1.0 E-13   \\
UGC~7878 &  323 &  53.05 &  (4)   & WFPC2/F814W    &   4.3 E-14      \\
UGC~7898 &  785 &  37.35 &  (6)   & WFPC2/F555W    &   $<$1.8 E-14   \\
UGC~8745 & --   &  --    &  --    & WFPC2/F814W    &       --        \\
UGC~9655 & --   &  --    &   --   & WFPC2/F702W    &       --        \\
UGC~9706 & 4009 &  30.79 &  (5)   & WFPC2/F702W    &   1.5 E-14      \\
UGC~9723 & 2879 &  34.18 &  (7)   & WFPC2/F814W    &         --	     \\
\hline	  		  	 			     
NGC~1316 & 2022 & 30.23  &  (8)   & NICMOS/F160W   &    $<$9.6 E-13  \\
NGC~1399 &  319 & 56.66  &  (4)   & WFPC2/F814W    &    1.4 E-14     \\
NGC~3258 & --   &    --  &  --    & ACS/F814W	   &      4.9 E-14   \\
NGC~3268 & --   &    --  &  --    & ACS/F814W	   &    $<$1.5 E-14  \\  
NGC~3557 & 3217 &  37.99 &  (1)   & WFPC2/F555W    &       --        \\
NGC~4373 & --   &    --  &   --   & WFPC2/F814W    &     2.4 E-14    \\ 
NGC~4696 & 1560 &  85.84 &  (9)   & ACS/F814W	   &     2.0 E-14    \\
NGC~5128 &463/1253 & 19.6+6.88 & (10)& NICMOS/F222W&     4.5 E-11    \\
NGC~5419 &4999/5000&15+15      & (1) & WFPC2/F555W &     6.7 E-14    \\ 
IC~1459  &  2196& 60.17  &  (11)  & WFPC2/F814W    &     4.4 E-13    \\
IC~4296  & 3394 & 25.4   &  (12)  & NICMOS/F160W   &     9.5 E-14    \\ 
IC~4931  & --   &  --    &  --    & WFPC2/F814 W   &     1.9 E-14    \\  
\hline						     
\end{tabular}

(1) this work, (2) \citet{balmaverde05}, (3) \citet{ho01b}, (4)
\citet{loewenstein01},
(5) \citet{filho04}, (6) \citet{randall04}, (7) \citet{terashima03}, (8) \citet{kim03},
(9) \citet{satyapal04}, (10) \citet{evans04}, (11) \citet{fabbiano03},
(12) \citet{pellegrini03}. 
\end{table*}

The basic
data for  the selected galaxies, namely the recession velocity  
(corrected for Local Group infall onto Virgo), the K
band  magnitude from  the Two  Micron  All Sky  Survey (2MASS),
the galactic extinction and the total and
core radio fluxes were given in \citetalias{capetti05}.
 
In the following three subsections, 
we derive and discuss the measurements for the
nuclear sources in the optical, X-ray and radio bands.

\subsection{Optical nuclei.}

\begin{figure*}
\centerline{
\psfig{figure=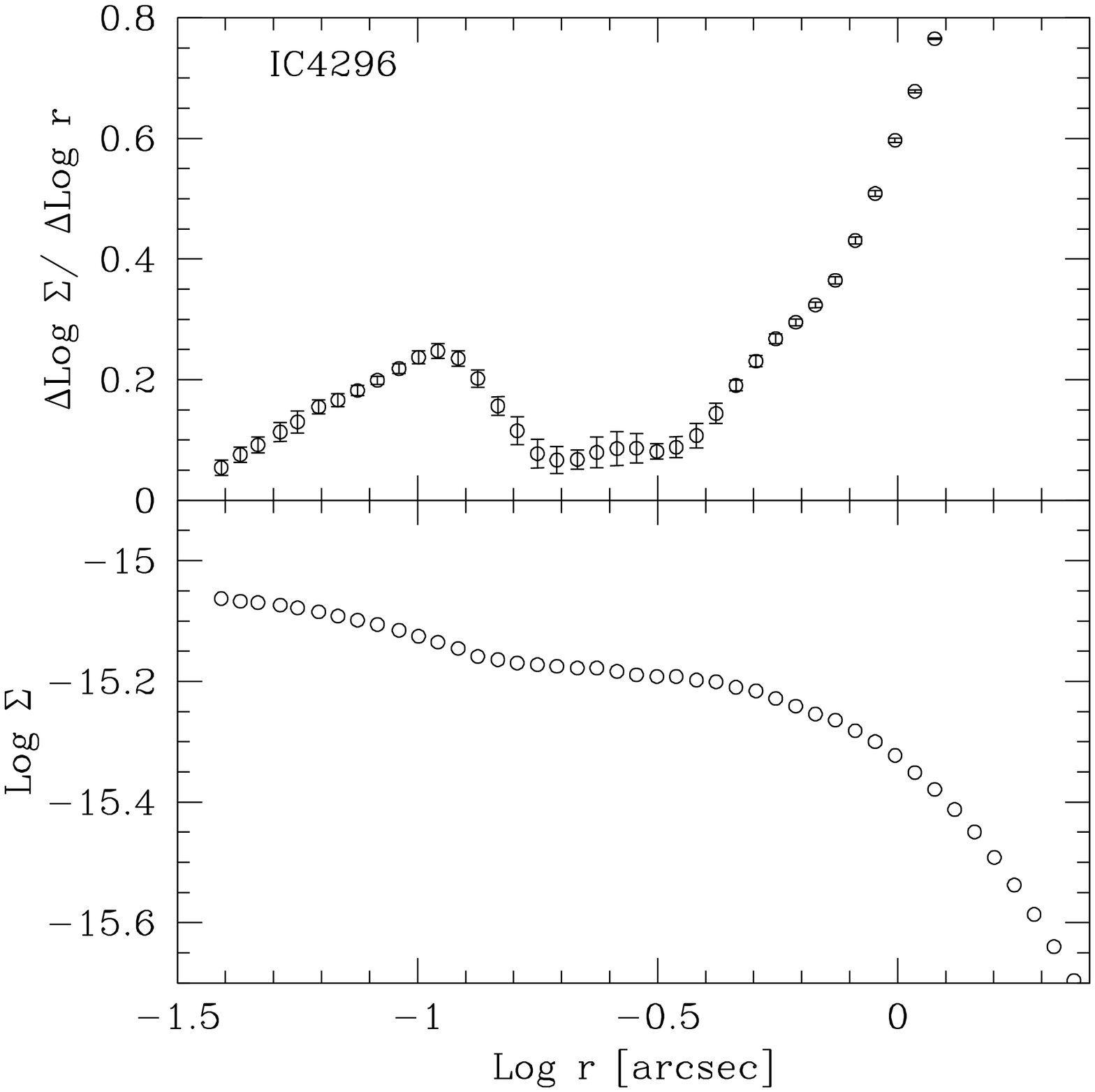,width=0.33\linewidth}
\psfig{figure=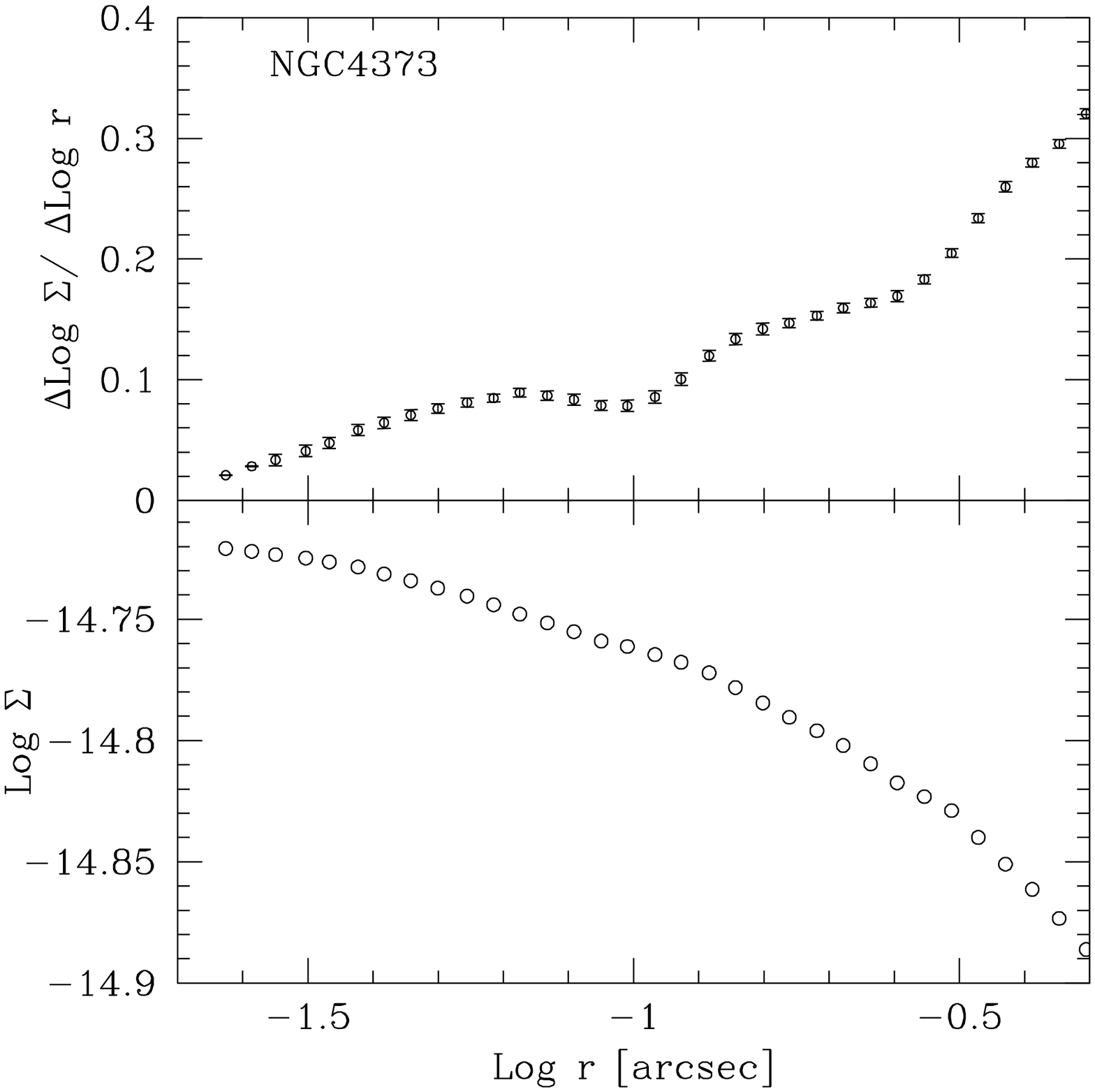,width=0.33\linewidth}
\psfig{figure=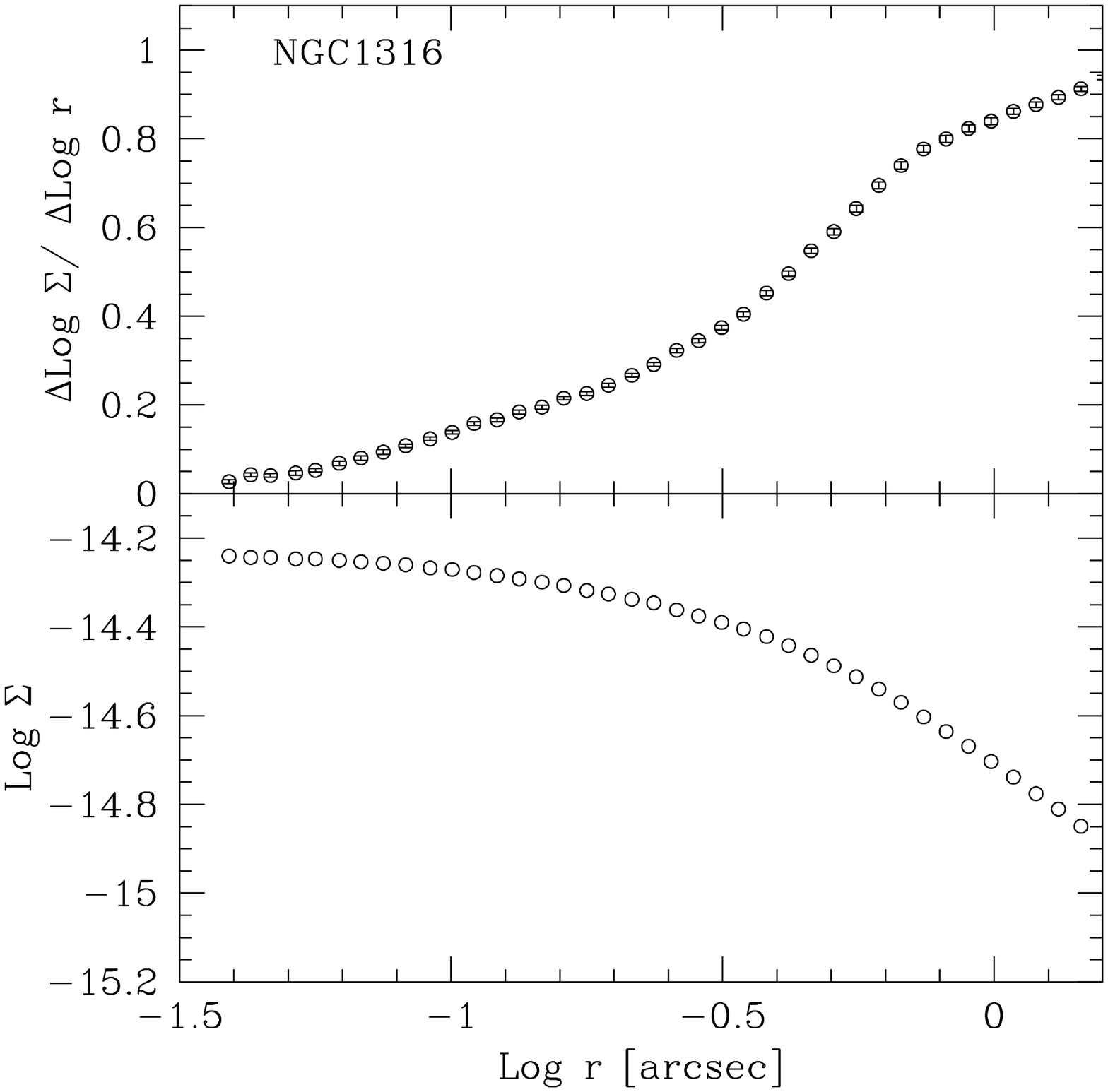,width=0.33\linewidth}
}
\caption{\label{hstnuc} Brightness profile and its derivative for 
three objects of the sample, namely IC~4296, NGC~4373 and NGC~1316.
The first two galaxies show, at decreasing significance level, the
characteristic up-turn in the profile associated with a nuclear source.
This is not seen in the third object which is then considered as a
non-detection.}
\end{figure*}

The detection and measurement of an optical nuclear source at the 
center of a galaxy is a challenging task particularly  
when it represents only a small 
contribution with respect to the host emission, 
as is likely often to be the case of the weakly active galaxies
making up our sample.

Different approaches have been employed in the literature.
The most widely used method is to fit the overall brightness 
profile of a galaxy
with an empirical functional form and to define a galaxy as
``nucleated'' when it shows a light excess in its central region
with respect to the model \citep[e.g.][]{lauer04,ravi01}.
The drawback of this ``global'' approach is that it 
assumes that the model can be extrapolated inwards
from the radial domain over which the fit was performed.
Furthermore, the measurement and identification of the nuclear
component are coupled with the behaviour of the brightness profile at all radii
and with the specific choice of an analytic form. 
Although this is not a significant issue for bright
point sources, it is particularly worrisome for the faint nuclei
we are dealing with.
Nonetheless, \citet{rest01} pointed out that in general nuclear light excesses
are associated with a steepening of the profile as the HST resolution
limit is approached. Indeed this is expected in the presence of
a nuclear point source, since the convolution with the Point Spread 
Function produces a smooth decrease of the slope toward the center
when only a diffuse galactic component is present.
We therefore preferred to adopt a ``local'' approach
to identify nuclear sources, based on the characteristic up-turn
they cause in the nuclear brightness profile.

More specifically, we  evaluated the  derivative  of  the
brightness profile in a log-log representation 
for the sources of our sample. In order to increase the 
stability of the slope measurement this has been estimated 
by combining the brightness 
measured over two adjacent points on each side of the radius of interest,
yielding a second order accuracy. We then look for  
the presence of a nuclear up-turn in the derivative 
requiring for a nuclear detection a difference larger than
3 $\sigma$ from the slopes at the local minimum and maximum.
This a rather conservative definition since 
the region over which the up-turn is detected 
extends over several pixels while we only consider the
peak-to-peak difference. 

To illustrate this we focus on three
cases. In the HST image as well as in the brightness profile of IC~4296 
a nucleus clearly stands out against the underlying background and
the central  steepening at about $r=0\farcs15$ is highly significant. 
NGC~4373 is the detection with the least significance of our sample,
in  which the presence of a  nucleus is uncertain
from just the visual inspection of 
the optical image,  but the
derivative of the brightness profile reveals the effect  of the point
source with an increase of 0.013 $\pm$ 0.004 from
$r=0\farcs1$ and $r=0\farcs07$.  
Instead  in NGC~1316 we do not  have evidence for any
compact point source, both in the image and in the brightness profile
derivative, and it is considered as a non-detection.

Adopting this strategy in 18 out of 29 objects we identify an optical nucleus,
with a  percentage of $\sim$ 60\% of the total sample. 
In seven objects we did not find any upturn and these are 
considered upper limits.
Note that this is again a conservative approach, since 
a nuclear source can still be
present but its intensity might not be sufficient to compensate the
downward trend of the derivative sets by the host galaxy. 

In the remaining 4 objects the central regions have a complex 
structure and no estimate
of the optical nucleus intensity can be obtained. 
In two cases (UGC~8745 and UGC~9723) 
the central regions are completely hidden by a kpc scale edge-on disk, 
while in NGC~3557 the study of its nuclear regions 
is hampered by the presence 
of a circumnuclear dusty disk. In UGC~9655, the innermost region
($r<0\farcs1$) has a lower brightness than its surrounding; since
only a single band image is available we cannot assess 
if this is due to dust absorption or
to a genuine central brightness minimum 
as in the cases discussed by \citet{lauer02}.

We measured the  nuclear luminosity with the task RADPROF  in IRAF,
choosing as the  extraction region a circle centered on the  nucleus with
radius set at the location of the up-turn and as the background region 
a circumnuclear annulus, 0.1\arcsec\ in width. 
For the undetected nuclei we set as upper limits the light excess
with respect to the starlight background
within a circular aperture 0.1\arcsec\ in diameter.
Then  we use  the PHOTFLAM and  EXPTIME keyword in  the image
header to convert  the total counts to fluxes. 
Errors on the measurements of the optical nuclei are dominated
by the uncertainty in the behaviour of the host's profile, 
while the statistical and absolute calibration errors amount
to less than 10 \%.h
The very presence of the nucleus prevents
us from determining accurately the host contribution within
the central aperture. Our strategy is to remove the background measured
as close as possible to the nucleus, i.e. effectively we adopted
a constant starlight distribution in the innermost regions. 
An alternative approach would be to extrapolate the profile with
a constant slope instead. Our definition of
nuclear sources (an increase in the profile's derivative) 
implicitly requires that the observed profile lies
above this extrapolation, but the resulting flux is reduced
by at most a factor of 2 (with respect to the case of constant
background) for the nuclei with the smallest contrast
against the galaxy. As will become clear in the next sections,
errors of this magnitude only have a marginal impact on our conclusions.
The resulting fluxes are reported
in Table \ref{tabsample2}.
We finally derived all the
luminosities referred to 8140 \AA\ (see Table \ref{lum}),
after correcting for the Galactic extinction as tabulated in 
\citetalias{capetti05} and
adopting an optical spectral 
index\footnote{We define the spectral index $\alpha$ with the spectrum
in the  form F$_\nu  \propto \nu^{-\alpha}$} $\alpha_o = 1$.

\subsection{X-ray nuclei}

For the measurements of the X-ray nuclei we concentrate 
only on the Chandra measurements, as this telescope provides 
the best  combination of
sensitivity and resolution necessary to detect the faint nuclei expected in
these weakly active galaxies.
Data for 21 core galaxies are available in the Chandra public archive.

When available, we used the results of the analysis of the X-ray
data from the literature. 
We find estimates of the luminosities of the nuclear sources (usually defined
as the detection of a high energy power-law component)
based on  Chandra data  for 16
objects of our sample (12 of which are detections and 4 are upper limits) 
which we rescaled to our adopted distance 
and converted to the 2-10 keV  band, using  the
published power law index. In
Tab \ref{tabsample2} we give a summary of the available
Chandra data, while references and details on the X-ray 
observations and analysis are presented in Appendix \ref{notes}. 

We also considered the Chandra archival data for the 5 unpublished objects,
namely UGC~5902, UGC~6297, UGC~7203, NGC~3557 and NGC~5419. We analyzed these
observations  using the Chandra  data analysis  CIAO v3.0.2,  with the
CALDB version 2.25, using the same strategy as in
\citet{balmaverde05}. 
We reprocessed all the data from level 1 to level 2,
subtracting the bad pixels,  applying ACIS CTI correction, coordinates
and pha randomization.  We searched for background flares and 
excluded some period of bad aspect.

We then extracted the spectrum in a circle region centered on the nucleus 
with a radius of 2\arcsec\ and we take the background in an annulus
of 4\arcsec. We grouped the spectrum to have at least 10 counts 
per bin and applied Poisson statistics. 
     
For two  objects  (NGC~3557 and NGC~5419) we obtain  a detection of a
nuclear power-law source by
fitting the spectrum using an absorbed power-law plus a thermal
model,  with  the  hydrogen  column  density  fixed  at  the  Galactic
value. Details of the results are given in Appendix \ref{notes}.
For the remaining 3 galaxies we 
set an  upper  limit to  any  nuclear  emission, with  the
conservative   hypothesis   that   all   flux  that   we   measure   is
non-thermal. We then fit the  spectrum with an  absorbed (to  the galactic
value)  power law  model with  photon index  $\Gamma=2$.  

The X-ray luminosities for all objects are given in Table \ref{lum}.

\subsection{Radio nuclei.}

The radio data available for all objects of our sample 
are drawn from the surveys by \citet{wrobel91b} 
and \citet{sadler89}, performed with the VLA at 5 GHz with a resolution of
$\sim$ 5\arcsec. Although these represent the most uniform
and comprehensive studies of radio emission in early-type galaxies,
they do not always have a resolution sufficient to separate the core
emission from any extended structure. \citet{sadler89} 
argued that at decreasing radio luminosity there is a corresponding
increase of the fractional contribution of the radio core.

To verify whether the VLA data overestimate the core flux, we searched the literature  for radio  core measurements  obtained at
higher  resolution  (and/or  higher  frequency) than  our  data.  This
would  improve  the  estimate  of the  core  flux  density,
avoiding the contribution of  extended emission or spurious sources to
the nuclear  flux as  well as revealing  any radio  structure.  Better
measurements, from  VLBI data  or from  higher  frequency/resolution VLA
data, are available for most CoreG  (23 out of 29) and compact cores
were detected in all but 2 objects. The radio core
fluxes are taken from \citet{nagar02} (15 GHz VLA data and 5 GHz VLBI
data), \citet{filho02} and \citet{krajnovic02} (8.4 GHz at the VLA),
\citet{jones97} (8.4 GHz VLBI data) and
\citet{slee94} (PTI 5 GHz interpolated data).

\begin{figure}
\centerline{
\psfig{figure=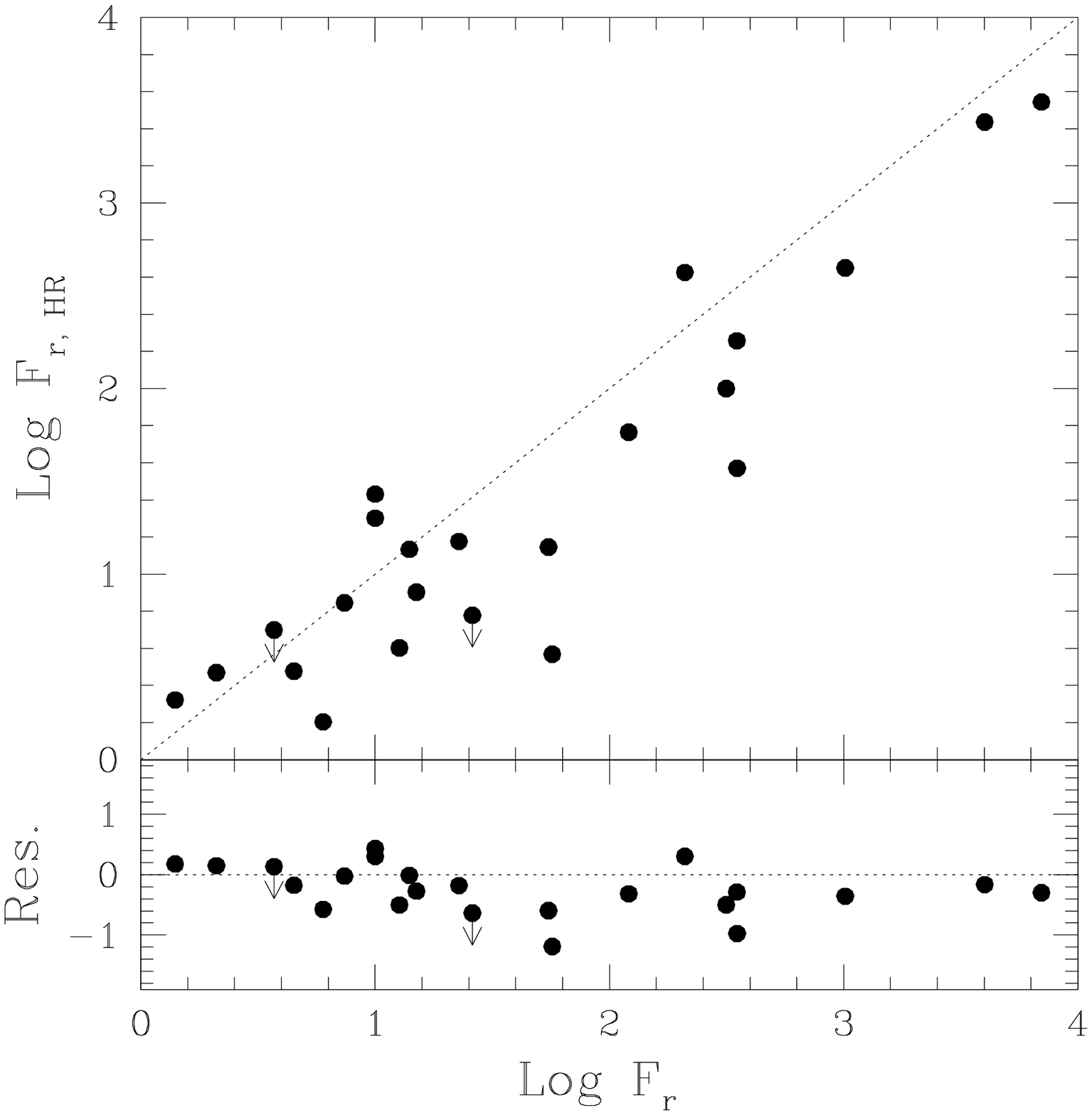,width=1.00\linewidth}
}
\caption{Radio core flux density for CoreG obtained at 
5 GHz with VLA (used in this work) compared to higher resolution data from
\citet{slee94,nagar02,filho02,
krajnovic02,jones97}.
The dotted line is the bisectrix of the plane.}
\label{cfr}
\end{figure}

In Fig. \ref{cfr} we
compare the radio  core flux density used in our  analysis 
against observations made at higher resolution.
Overall there  is a  substantial agreement between the two datasets,
with a  median difference  of only
$\sim$0.25 dex (a factor  $\sim$1.6), with only two objects
substantially offset (by a factor of $\sim$ 10).
However, since  these data are  highly inhomogeneous and  given the
general  agreement with the  5 GHz VLA measurements, we
prefer to retain the values of \citeauthor{wrobel91b} and \citeauthor{sadler89}.
Nonetheless, we always checked that using these higher resolution 
core fluxes our main results are not significantly affected 
(see Appendix \ref{radionuc} for a specific example).

\section{The multiwavelength properties of nuclei of core galaxies.}
\label{nuclei}

\begin{table*}
\caption{Core galaxies data: (1) UGC name, (2) intrinsic nuclear X-ray luminosity (2-10 keV) [erg s$^{-1}$], (3) nuclear optical luminosity (8140 \AA)[erg s$^{-1}$] 
corrected for absorption using the galactic extinction values in Paper I,
(4) nuclear radio luminosity (5GHz) [erg s$^{-1}$] derived from Paper I, (5) total radio luminosity (5GHz) [erg s$^{-1}$] derived from Paper I, 
(6) H$_{\alpha}$+[NII] line luminosity [erg s$^{-1}$] from Ho et al. 1997 or $^{a}$ Phillips et al. 1986,
 (7) total K band galaxy's absolute magnitude from 2MASS, (8) logarithm of black hole mass in solar unity 
from $^{b}$ Marconi et al. 2003 or derived using the velocity dispertion.}
\label{lum}
\centering
\begin{tabular}{l c c c c c c c c}
\hline \hline
Name     & Log L$_{x}$ &   Log $\nu$ L$_{o}$  &  Log $\nu$ L$_{core}$  &   Log $\nu$ L$_{tot}$ &  Log $L_{H_{\alpha}+[NII]}$  & $M_{K}$ &  Log $(M_{BH}/M_{\odot})$  \\
\hline  
UGC~0968 &       --   &   $<$39.77 &    36.94   &    36.94   &    38.98   & -25.39   &  8.54        \\
UGC~5902 &  $<$38.40  &  $<$39.10  &    35.83   &    35.83   &    38.76   & -24.25   &  8.00$^{b}$  \\
UGC~6297 &  $<$38.40  &  $<$39.06  &    36.46   &    36.46   &    39.09   & -23.40   & 	  8.33	    \\
UGC~7203 &  $<$38.93  &    39.85   &    37.44   &    37.44   &    38.74   & -24.08   &  7.98	    \\
UGC~7360 &    40.95   &    39.71   &    39.22   &    40.64   &    39.76   & -25.11   &  8.72$^{b}$  \\
UGC~7386 &    39.72   &    38.76   &    38.38   &    38.38   &    39.60   & -22.97   &  8.43	    \\
UGC~7494 &     39.30  &    39.27   &    38.57   &    39.48   &    39.14   & -24.41   &  9.00$^{b}$  \\
UGC~7629 &    38.23   &    38.79   &    37.73   &    37.95   &    37.98   & -25.09   &  8.78	    \\
UGC~7654 &     40.30  &    40.72   &     39.90  &    41.15   &    40.00   & -25.48   &  9.53$^{b}$  \\
UGC~7760 &     38.40  &    38.73   &     37.30  &     37.3   &    37.90    & -21.86   &  8.54	    \\
UGC~7797 &       --   & $<$40.19   &    38.05   &    38.05   &    39.46   & -24.61   &  8.33	    \\
UGC~7878 & $<$38.41   &    39.07   &     36.90  &    37.77   &    38.60   & -24.43   &  8.16	    \\
UGC~7898 & $<$38.52   & $<$39.13   &    37.46   &    37.59   &      --    & -25.34   &  9.30$^{b}$  \\
UGC~8745 &       --   &   Dusty    &    37.81   &    37.97   &    39.21    & -25.07   &  8.39	    \\
UGC~9655 &       --   &   Dusty    &    36.96   &    36.96   &    39.01    & -24.74   &  8.44	    \\
UGC~9706 &    38.26   &    39.25   &    37.31   &    37.31   &    39.11   & -25.06   &  8.43	    \\
UGC~9723 &  $<$38.18  &   Dusty    &     36.92  &     36.92  &    38.28    & -23.82   &  7.73	    \\
\hline	  	       		    		 	      		     	
NGC~1316 &    39.62   &  $<$40.11  &    37.82   &    41.24   &      --        &  -25.99 &   8.36    \\
NGC~1399 & $<$38.79   &    38.65   &     37.20  &    38.73   &      --        &  -24.75 &   9.07    \\
NGC~3258 &       --   &    39.91   &    37.42   &    38.57   &      --        &  -24.39 &   8.67    \\
NGC~3268 &       --   & $<$39.42   &    38.21   &    38.21   &    39.67$^{a}$ &  -24.55 &   8.33    \\
NGC~3557 &    40.08   &    Dusty   &    37.94   &    39.41   &      --        &  -25.70 &   8.67    \\
NGC~4373 &       --   &    39.79   &    38.15   &    38.15   &      --        &  -25.51 &   8.49       \\
NGC~4696 &    40.04   &    39.59   &    38.65   &    40.05   &    39.30       &  -25.69 &   8.55       \\
NGC~5128 &    42.11   &     40.31  &    39.05   &    40.31   &    38.16       &  -24.64 &   8.38$^{b}$ \\
NGC~5419 &    40.69   &     40.79  &    38.42   &    39.83   &    --          &  -26.14 &   9.02	\\
IC~1459  &    40.56   &    40.33   &    39.38   &    39.38   &    40.02$^{a}$ &  -24.70 &   9.18$^{b}$  \\
IC~4296  &    41.18   &    39.85   &    39.46   &    40.35   &    39.74$^{a}$ &  -25.91 &   9.04	\\
IC~4931  &       --   &    40.19   &    37.51   &    37.51   &      --        &  -25.79 &   8.67        \\
\hline
\end{tabular}
\end{table*}

\begin{figure*}
\centerline{
\psfig{figure=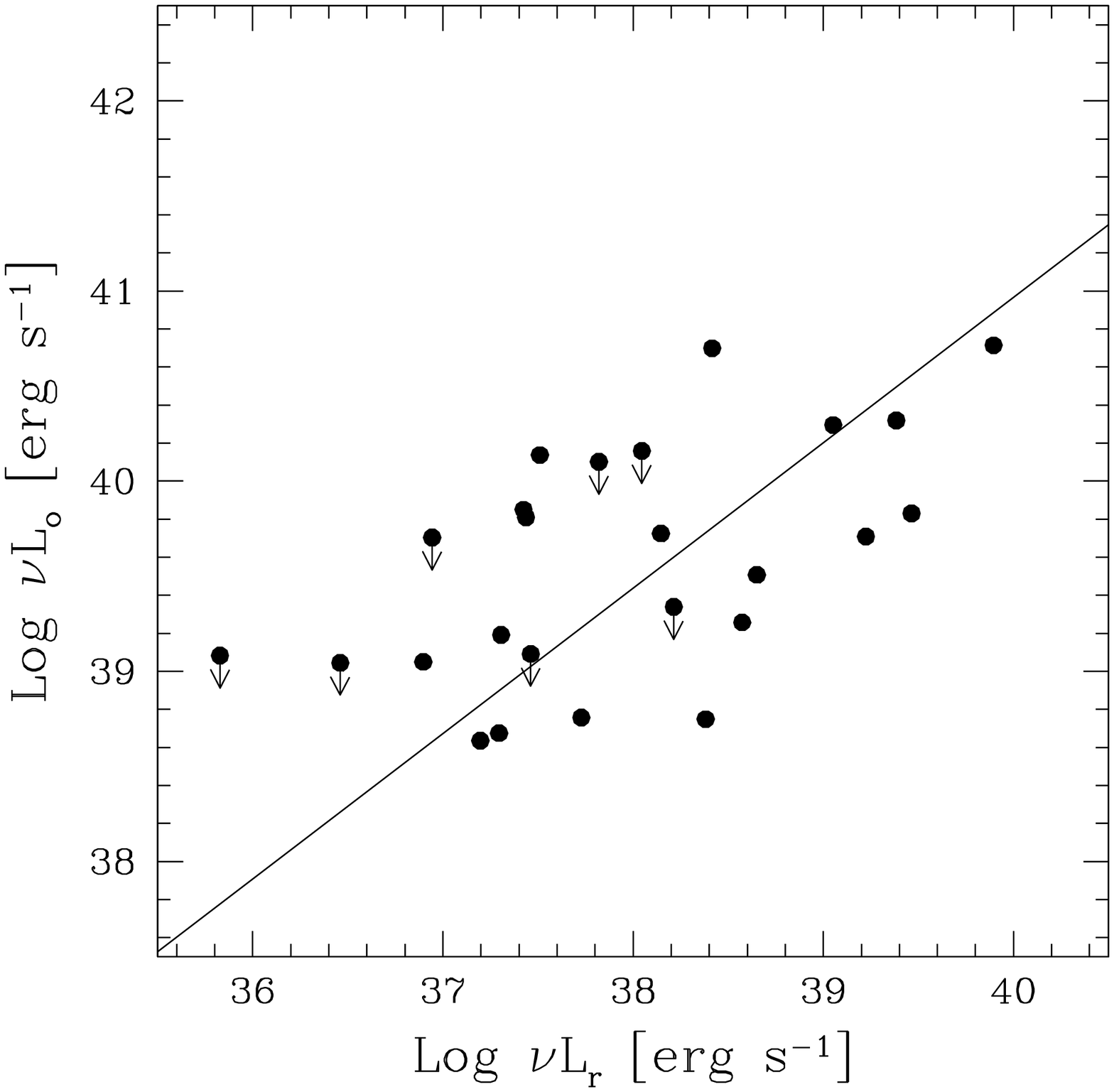,width=0.50\linewidth}
\psfig{figure=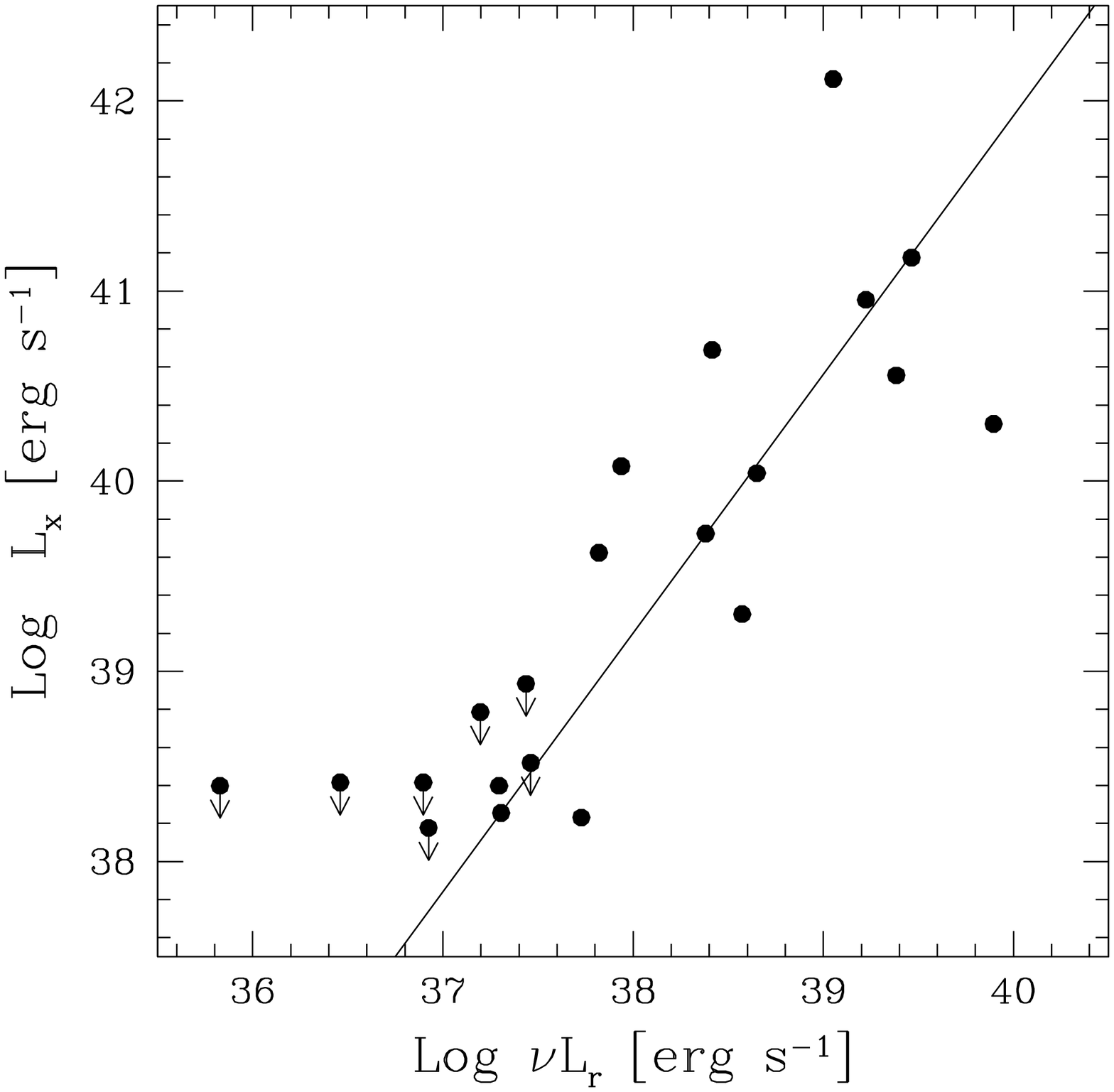,width=0.50\linewidth}
}
\caption{\label{corr1} Radio core luminosity for the early-type galaxies
with a ``core'' profile versus the optical (left) 
and X-ray (right) nuclear luminosities.}
\end{figure*}

Having collected the multiwavelength  information for the nuclei of our
core  galaxies we  can compare  the emission  in the  different bands.
First of all, we can estimate the ratio between the radio, optical and
X-ray  luminosities: the  median values  are Log$(\nu L_r/\nu L_o)  \sim -1.5$
(equivalent to  a standard radio-loudness parameter Log  R $\sim$ 3.6)
\footnote{R =  L$_{5\rm {GHz}}$ / L$_{\rm B}$. As in
  Sect. \ref{nuc} we transformed the optical fluxes to the B band
adopting an optical spectral index $\alpha_o = 1$.}
and Log R$_{\rm X}$  = Log$(\nu L_{\rm r}/L_{\rm X})  \sim -1.3$,  
both with a  dispersion of  $\sim$ 0.5
dex.  These  ratios are clearly  indicative of a radio-loud  nature for
these  nuclei when compared  to both  the traditional  separation into
radio-loud  and radio-quiet  AGN 
\citep[Log R =  1, e.g.][]{kellermann94}, as well  as with  the
radio-loudness threshold introduced by \citet{terashima03}
based on  the X-ray to radio luminosity  ratio  (Log R$_{\rm X}$ = -4.5).  
Furthermore, the nuclear
luminosities  in   all  three   bands  are clearly correlated (see
Fig. \ref{corr1} and Table \ref{tab0} for a summary of the results of
the statistical analysis): the generalized (including the presence
of upper limits) Spearman rank correlation coefficient
$\rho$ is 0.63 and 0.89 for $L_{\rm r}$ vs. $L_{\rm o}$ 
and $L_{\rm r}$ vs. $L_{\rm X}$ respectively, with probabilities that
the correlations are not present of only 0.002 and 0.0001.

\begin{figure*}
\centerline{
\psfig{figure=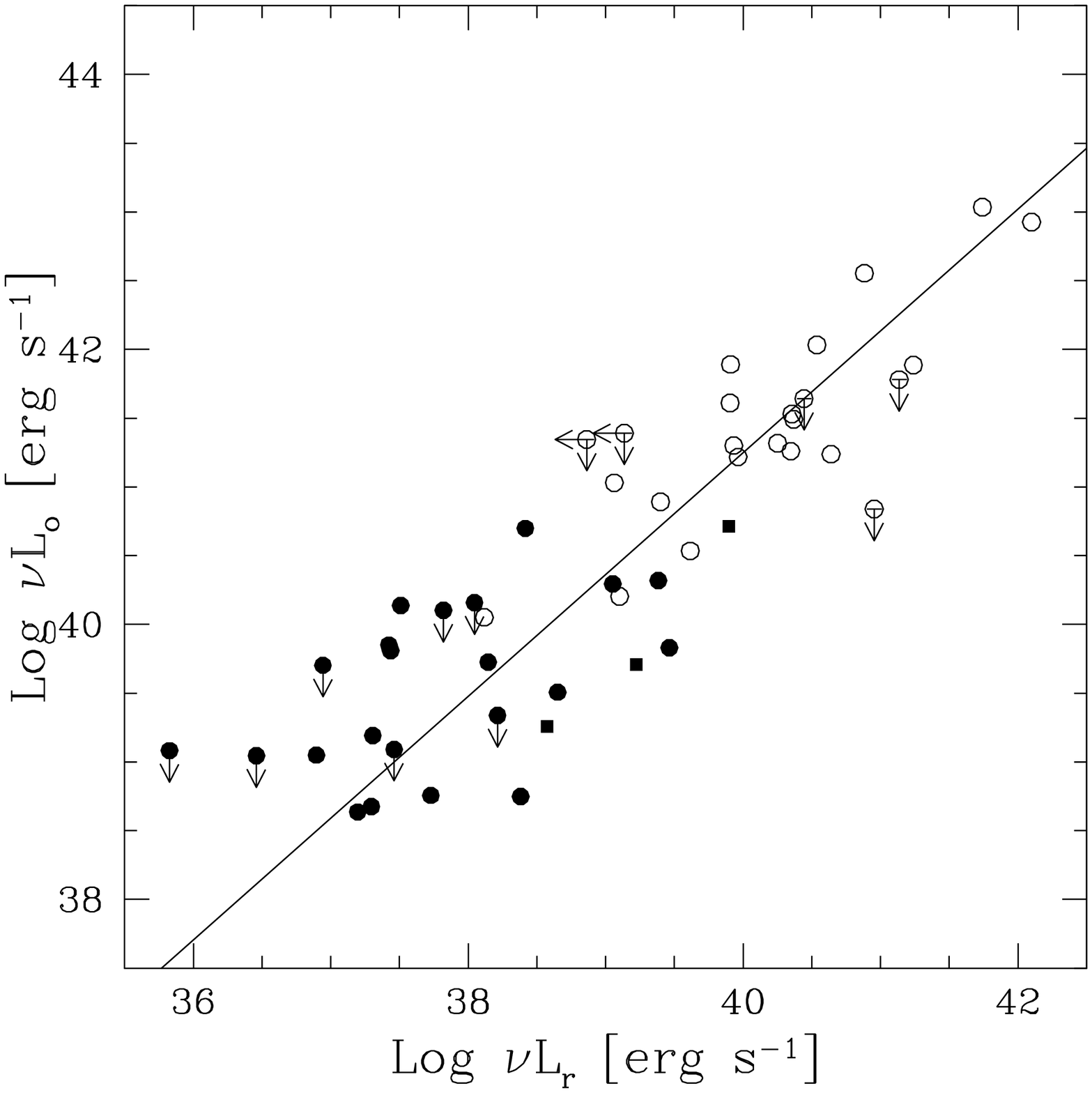,width=0.50\linewidth}
\psfig{figure=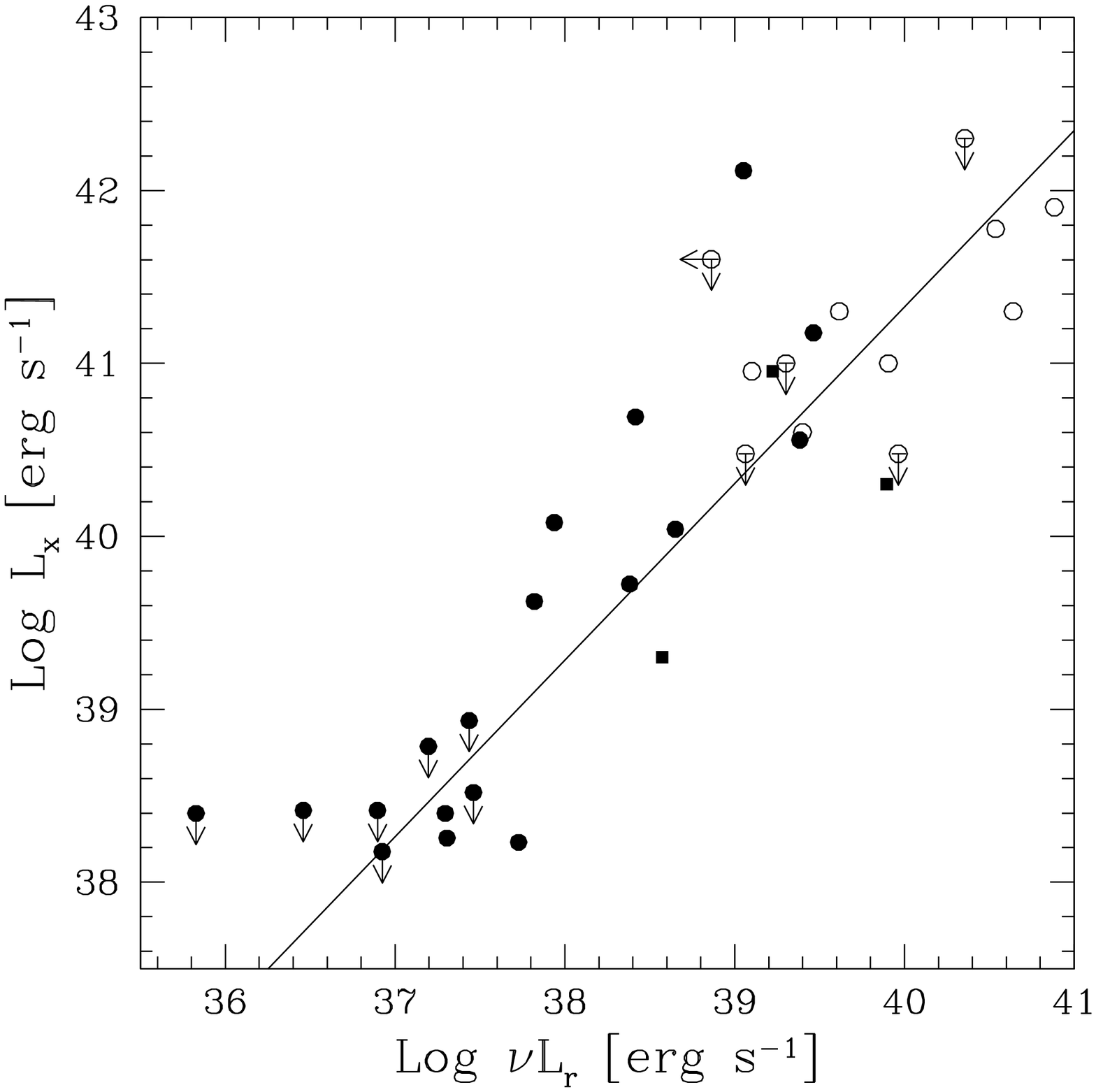,width=0.50\linewidth}
}
\caption{\label{corr} Comparison of radio and optical 
(left) and X-ray (right) nuclear luminosity
for the sample of core-galaxies (filled circles) and for the 
reference 3C/FR~I sample of low luminosity radio-galaxies (empty
circles). 
The three sources in common are marked
with a filled square. The solid lines reproduce the best linear fits.
}
\end{figure*}

Both results are reminiscent of what is  observed for the
radio-loud nuclei 
of low luminosity radio-galaxies (LLRG). \citet{chiaberge:ccc} 
and \citet{balmaverde05} reported on similar multiwavelength
luminosity trends for the sample of
LLRG formed  by the  3C sources with  FR~I morphology.
The connection between the CoreG and LLRG becomes 
more evident if we add LLRG  in the diagnostic
planes (see  Fig. \ref{corr} and Table \ref{lumfr1}). 
The early-type core  galaxies 
follow the same behaviour of the stronger radio galaxies,
extending it downward by 3 orders  of magnitude in radio-core
luminosity as they reach levels
as low as $L_{\rm r} \sim 10^{36}$ erg s$^{-1}$.

We estimated the best linear fit for the combined CoreG/LLRG sample
in both the 
$L_{\rm r}$ vs. $L_{\rm o}$ and $L_{\rm r}$ vs. $L_{\rm X}$ planes. 
The best  fits were  derived  as the
bisectrix of the  linear fits using the two  quantities as independent
variables following the suggestion by \citet{isobe90} that this is
preferable for problems needing symmetrical treatment of the variables. 
The presence of upper limits in the independent variable
suggests that we could take advantage of the 
methods of survival analysis proposed by e.g. \citet{schmitt85}.
However, the drawbacks discussed by \citet{sadler89}
and, in our specific case, 
the non-random distribution of upper limits, argue against this approach. 
We therefore preferred to exclude  upper limits from the  linear
regression analysis. Nonetheless, a posteriori, 
1) the  objects  with an undetected nuclear
component in  the optical or X-ray are
consistent  with the correlation  defined by  the detections  only;
2) the application of the Schmidt methods provides correlation
slopes that agree, within the errors, with our estimates.
 
We obtained (indicating the Pearson 
correlation coefficient with $r$ and slope with $m$)
$r_{ro}$=0.90 and $m_{ro}=0.89  \pm 0.07 $, 
$r_{rx}$=0.89 $m_{rx} = 1.02\pm 0.10 $ for the radio/optical and
radio/X-ray correlations respectively.
The slopes and normalizations derived for CoreG, LLRG and the
combined CoreG+LLRG sample 
(see Table \ref{tab0}) are consistent within the errors
and this indicates that there is no significant change in the
behaviour between the two samples. 
Only the dispersion is slightly larger for the CoreG nuclei 
being a factor of $\sim 4$ rather than $\sim 2$ for the LLRG sample alone. 

\citet{chiaberge:ccc} first reported the presence of a correlation
between radio and optical emission in the LLRG and they concluded that
this is most likely due to a common non-thermal jet origin for the
radio and optical cores. 
Recently \citet{balmaverde05} extended the analysis to the X-ray cores;
the nuclear X-ray luminosity also correlates with those of the radio
cores and with a much smaller dispersion 
($\sim$ 0.3 dex) when compared to similar
trends found for other classes of AGN \citep[see e.g.][]{falcke95}, 
again pointing to a common origin for the emission in the three bands.
Furthermore, the broad band spectral indices of the 3C/FR~I cores 
are very similar to those measured in BL Lacs objects (for which a jet
origin is well established) in accord with the
FR~I/BL~Lacs unified model 
(we will return to this issue in Section \ref{bllac}).

The core galaxies of our sample thus appear to smoothly extend the results
obtained for LLRG to much lower radio luminosity, expanding the
multiwavelenght nuclear correlations to a total of 6 orders of
magnitude. This strongly argues in favour of a jet
origin for the nuclear emission also in the core galaxies and that they 
simply represent the scaled down versions of these already low luminosity AGN.

\begin{table}
\caption{Correlations summary}
\begin{tabular}{l l c c c l } \hline \hline
Sample  & Var. A    & Var. B & r$_{AB}$& Slope & rms     \\
\hline                                       
CoreG     & L$_{O}$ & L$_{r}$   & 0.59 & 0.76$\pm$0.21  & 0.62  \\
          & L$_{X}$ & L$_{r}$   & 0.78 & 1.36$\pm$0.20  & 0.59  \\
LLRG      & L$_{O}$ & L$_{r}$   & 0.94 & 0.82$\pm$0.11  & 0.32  \\
          & L$_{X}$ & L$_{r}$   & 0.95 & 0.99$\pm$0.11  & 0.33  \\
LLRG+CoreG& L$_{O}$ & L$_{r}$   & 0.90 & 0.89$\pm$0.07  & 0.56  \\
          & L$_{X}$ & L$_{r}$   & 0.89 & 1.02$\pm$0.10  & 0.58  \\
\hline
\end{tabular}
\label{tab0}
\end{table}

\subsection{Core galaxies vs. low luminosity radio-galaxies.}
\label{cg-fri}

\begin{table*}
\caption{Radio galaxies of the 3C/FR~I sample data: (1) name, (2) intrinsic nuclear X-ray luminosity (2-10 keV) [erg s$^{-1}$], 
(3) nuclear optical luminosity (8140 \AA)[erg s$^{-1}$], (4) nuclear radio luminosity (5GHz)[erg s$^{-1}$] and 
(5) total radio luminosity (178MHz)[erg s$^{-1}$] from Chiaberge et al. 1999,
 (6) H$_{\alpha}$+[NII] line luminosity from Capetti et al. 2005 [erg s$^{-1}$], (7) total K band galaxy's absolute magnitude from 2MASS, 
(8) logarithm of black hole mass in solar unity derived using the velocity dispertion or from $^{a}$ Marconi et al. 2003.}
\label{lumfr1}
\centering
\begin{tabular}{l c c c c c c c}
\hline \hline
Name     & Log L$_{x}$ &   Log $\nu$ L$_{o}$  &  Log $\nu$ L$_{core}$  &   Log $\nu$ L$_{tot}$ &  Log $L_{H_{\alpha}+[NII]}$  & $M_{K}$  &  Log $(M_{BH}/M_{\odot})$  \\
\hline  
         3C~028    & $<$41.60  &  $<$41.35 &  $<$38.86  &     42.32  &       --   & -26.06    &    --       \\
         3C~029    &      --   &    41.32  &     40.25  &     41.01  &     40.40  & -25.72    &    -8.11     \\
         3C~031    &    40.60  &    40.89  &      39.40 &     40.21  &     39.89  & -25.67    &   -8.70     \\
         3C~066B   &   41.00   &    41.61  &      39.90 &     40.59  &     40.22  &  --       &    --       \\
         3C~075    & $<$ 41.00 &      --   &      39.30 &     40.28  &       --   &  --       &    8.84     \\
         3C~076.1  &      --   &      --   &       --   &     40.64  &      --    &  --       &    8.88	    \\
	 3C~078    &    41.90  &    42.55  &     40.88  &     40.74  &     40.71  &  -26.24   &    8.61	    \\
         3C~083.1  &   40.95   &     40.2  &      39.10 &     40.75  &     39.26  &  --       &    --	    \\
         3C~084    &    42.60  &    42.93  &      42.10 &     40.63  &       --   &  -26.11   &    8.58     \\
         3C~089    &      --   & $<$40.84  &     40.95  &     42.12  &       --   &  -26.05   &    8.85	    \\
	 3C~189    &   41.78   &    42.03  &     40.54  &     42.80  &       --   &  --       &    --	    \\
         3C~264    &      --   &    41.89  &     39.91  &     40.57  &     40.02  &  -25.09   &    8.67	    \\
	 3C~270    &    40.95  &     39.71 &      39.22 &     41.27  &     39.54  &  -25.11   &    8.72$^{a}$    \\
	 3C~272.1  &	39.30  &    39.26  &    38.57   &     40.10  &     38.56  &  -24.41   &    9.00$^{a}$    \\
	 3C~274	   &	40.30  &    40.71  &     39.90  &     41.78  &     39.15  &  -25.48   &    9.53$^{a}$    \\
	 3C~277.3  &      --   &     41.30 &     39.93  &     41.35  &     40.51  &  -25.20   &    --	     \\
	 3C~288    &      --   &    41.89  &     41.24  &     42.59  &       --   &   --      &     --       \\
	 3C~293    &      --   &      --   &     40.29  &     40.94  &       --   &   -25.44  &     8.14     \\
	 3C~296    &    41.30  &    40.53  &     39.62  &     40.39  &       --   &   --      &     8.80     \\
         3C~305    &      --   &      --   &     39.68  &     40.96  &       --   &   -25.45  &     8.07     \\
	 3C~310    &      --   &    41.26  &     40.35  &     41.74  &       --   &   --      &     8.21     \\
	 3C~314.1  &      --   & $<$41.39  &  $<$39.14  &     41.71  &       --   &   --      &     --	     \\
	 3C~315    &      --   &      --   &     41.23  &     41.85  &       --   &   --      &     --	     \\
	 3C~317    &    41.30  &    41.24  &     40.64  &     41.27  &       --   &   -26.13  &     8.32     \\
         3C~338    & $<$ 40.48 &    41.22  &     39.96  &     41.16  &     40.54  &   --      &     8.92     \\
         3C~346    &    43.30  &    43.03  &     41.74  &     41.99  &     41.41  &   -26.32  &      --	     \\
         3C~348    & $<$42.30  &    41.53  &     40.36  &     43.45  &       --   &   -26.40  &      --	     \\
         3C~424    &      --   & $<$41.64  &     40.44  &     41.88  &       --   &   --      &      --      \\
	 3C~433    &      --   &      --   &     39.69  &     42.29  &       --   &   --      &      --	     \\
	 3C~438    & $<$42.60  & $<$41.78  &     41.14  &     43.12  &       --   &   -27.00  &      --	     \\
	 3C~442    &      --   &    40.05  &     38.12  &     40.57  &     39.72  &   --      &       --     \\
	 3C~449    & $<$40.48  &    41.03  &     39.06  &     40.11  &     39.50  &   --      &      8.54    \\
	 3C~465    &      --   &    41.49  &     40.37  &     41.06  &     40.72  &   --      &      9.14     \\
\hline		   										 
\end{tabular}	   										
\end{table*}

The results presented above indicate that the nuclei of the CoreG
show a very similar behaviour to those of LLRG.
Here we explore in more detail how CoreG and LLRG
compare in their other properties, such
as the structure of the host, black hole mass, radio-morphology
and optical spectra.

Our sample was selected to include only
early-type galaxies with a core profile, e.g. with an asymptotic slope 
(toward the nucleus) of their surface brightness profiles $\gamma < 0.3$. 
Recently \citet{deruiter05} showed, from the analysis of a combined
sample of B2 and 3C sources, that they are all hosted
by early-type galaxies and that the presence of a flat core
is a characteristic of the host galaxies of all nearby radio-galaxies.

\begin{figure*}
\centerline{
\psfig{figure=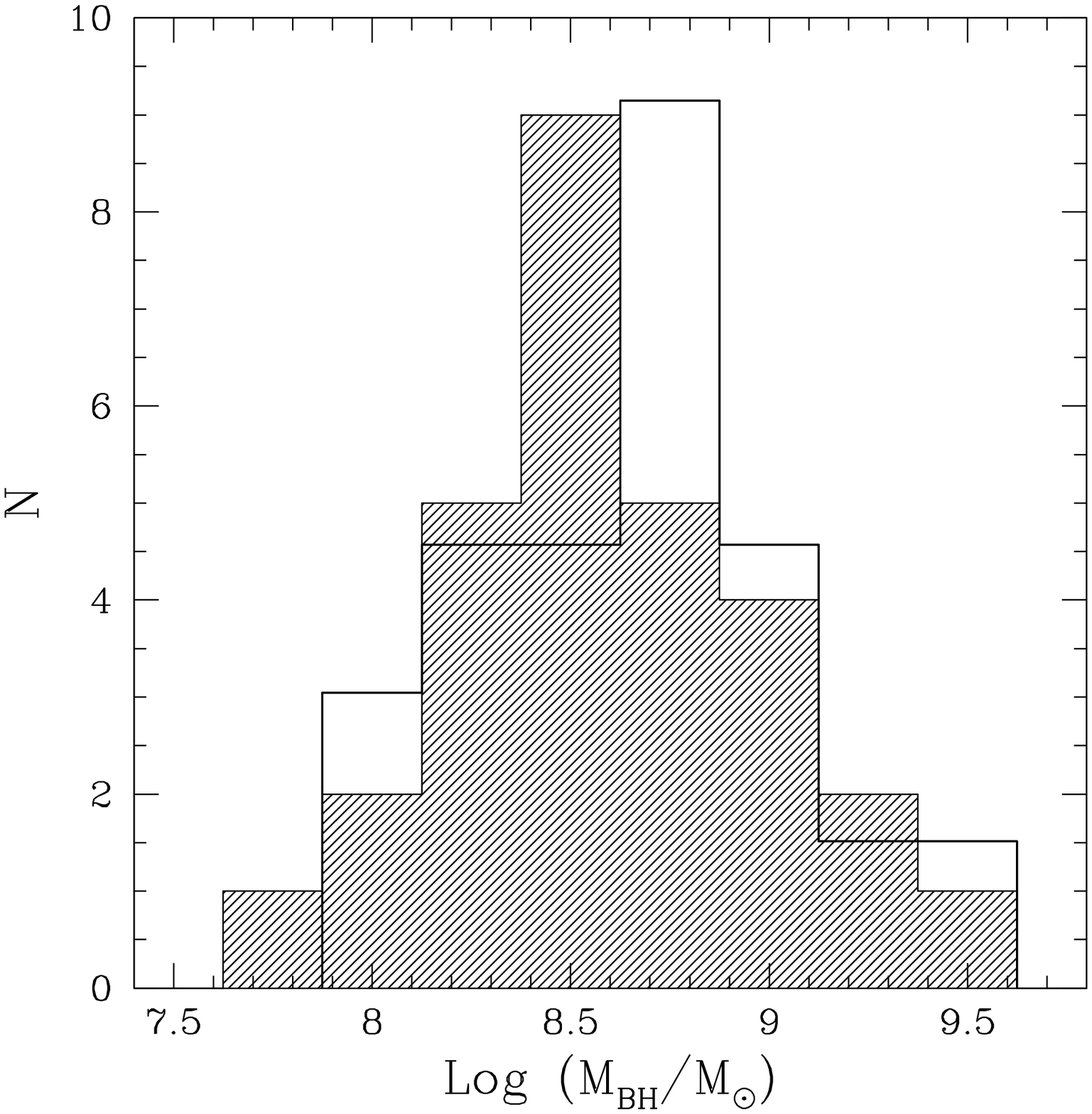,width=0.5\linewidth}
\psfig{figure=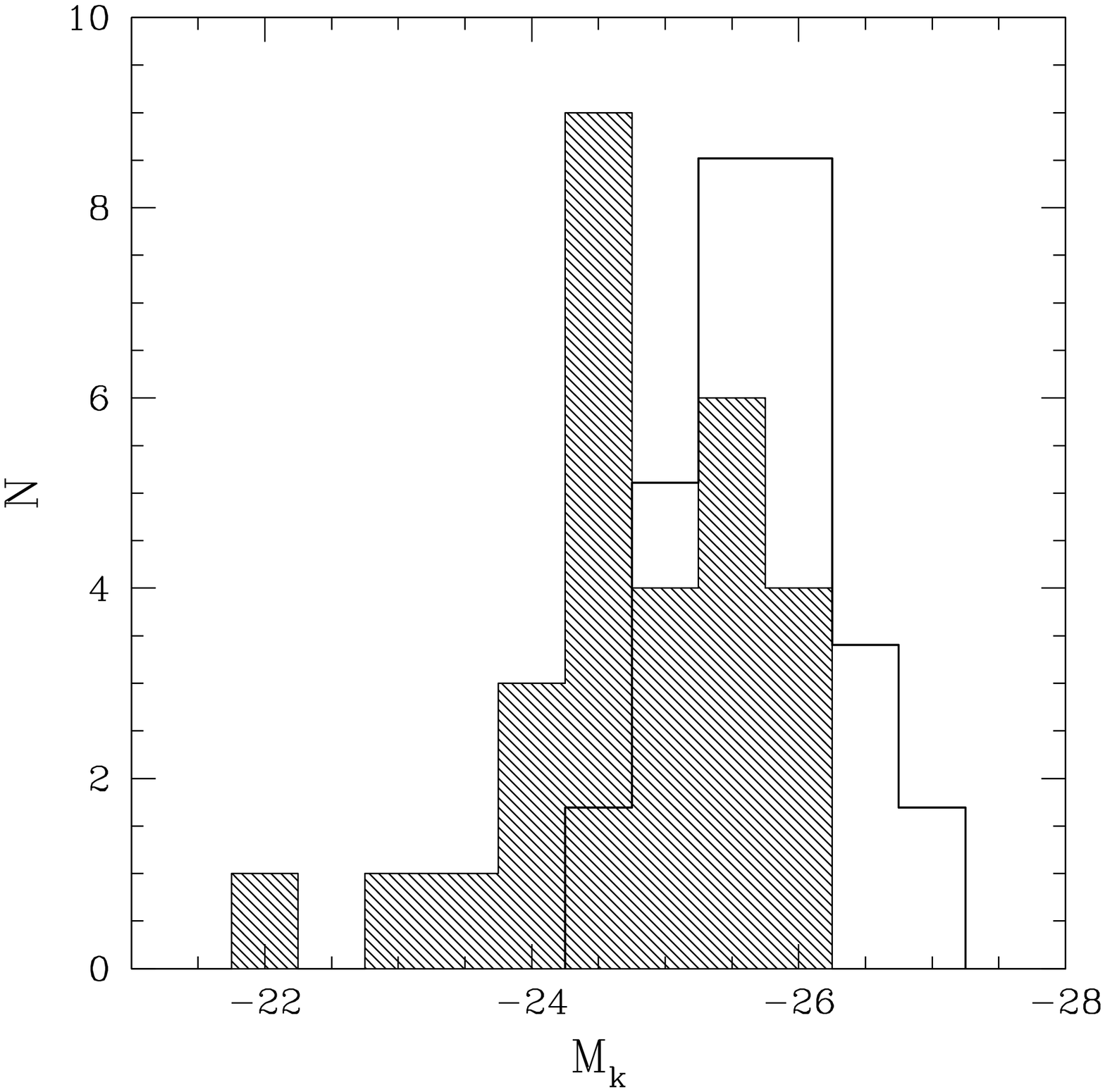,width=0.5\linewidth}
}
\caption{\label{mbhhis} Distributions for CoreG (shaded histograms) 
and for LLRG (empty histogram) of 
(left panel) black hole mass M$_{BH}$ and 
(right panel) absolute magnitude M$_K$.  
The LLRG histograms have been re-normalized multiplying 
by a factor 29/19 for M$_{BH}$ and 29/17 for M$_K$ respectively, 
i.e. the number of objects
in the two samples for which estimates of these parameters are available.}
\end{figure*}

A strong similarity between CoreG and LLRG emerges when 
comparing the mass of their
supermassive black holes. 
When no direct measurement 
\citep[taken from the compilation by][]{marconi03} 
was available, we estimated $M_{BH}$ using
the relationship with the stellar velocity dispersion
(taken from the LEDA database) 
in the form given by \citet{tremaine02}.
The distributions of $M_{BH}$ (see Fig. \ref{mbhhis}) of
the two samples are almost indistinguishable\footnote{The 
probability that the two samples are drawn 
from the same parent distribution is 0.32, 
according to the Kolmogorov-Smirnoff test.}, as they 
have median values of Log $M_{BH} = 8.54 $ and Log $M_{BH} = 8.70 $,
for CoreG and LLRG respectively,
and they also cover  
the same range, with most objects with Log $M_{BH} = 8 - 9.5$.

Further indications  of the  nature of CoreG  cores and  their connection
with LLRG come from  the emission lines in their optical spectra.
LLRG are characterized as a class by their LINER spectra
\citep[e.g.][]{lewis03} and this
is the case also for the CoreG of our sample.
In the NED  database, although about half of the CoreG do not have
a spectral  classification, 13 objects are classified 
as LINERs\footnote{This result provides further support to the 
suggestion by \citet{chiaberge05}
that a dual population is associated with galaxies with a LINER
spectrum, being formed by both radio-quiet and by radio-loud objects.
The CoreG are part of this latter sub-population of radio-loud LINER.}.  
The only exception is UGC~7203, with a Seyfert spectrum, but its
diagnostic line  ratios are borderline  with those of LINERs \citep{ho97}.
Concerning the emission line luminosity, \citet{capetti:cccriga} found a
tight relationship  between radio core  and line luminosity  studying a
group of LLRG formed by the 3C/FR~I complemented by
the sample of 21 radio-bright ($F_r > 150$ mJy) 
UGC galaxies defined by \citet{noelstorr03}. Line luminosity for 
our CoreG clearly follow the same trend defined by LLRG, 
although with a substantially larger  dispersion, not unexpected given their
low line luminosity and the non uniformity of the data used for this
analysis. 

Considering the radio structure,
several objects of our CoreG sample have a radio-morphology with well developed 
jets and lobes: UGC~7360, UGC~7494 and UGC~7654 are FR~I
radio-galaxies part of the 3C sample (3C~270, 3C~272.1 and 3C~274), 
while in the Southern sample we
have the well studied radio-galaxies NGC~1316 (Fornax A), a FR~II
source, NGC~5128 (Cen A) and IC~4296.
A literature search shows that at least another 11 sources 
have extended radio-structure indicative of a collimated outflow,
although in several cases this can only be seen in high
resolution VLBI images, such as
the mas scale double-lobes in UGC~7760 or the one-sided jet of 
UGC~7386 \citep{nagar02,falcke00}. 

Conversely, hosts of 3C/FR~I radio-sources are on average more 
luminous than core-galaxies
(see Fig. \ref{mbhhis}, left panel) although there is
a substantial overlap between the two groups: the median values are
$M_K=-24.8$ and $M_K=-25.7$ for CoreG and 3C/FR~I respectively,
with a KS probability of only 0.003 of being drawn from the same population. 
This reflects the well known trend,
already noted by \citet{auriemma77}, for which 
a brighter galaxy has a higher probability of being a stronger
radio emitter, and which is present also in our sample \citepalias{capetti05}. 
The selection of 
relatively bright radio sources, such as the 3C/FR~I, corresponds
to a bias toward more luminous galaxies. Indeed, within our sample,
imposing a threshold in total radio-luminosity of 
$L_{\rm tot} > 10^{39}$ erg/s,\footnote{The 5 GHz luminosity was
converted to 178 MHz for consistency with the 3C/FR~I values adopting
a spectral index of 0.7.} the low end for LLRG,  
decreases the median magnitude to -25.1,
in closer agreement with the 3C/FR~I value.

We conclude that the properties of our low radio luminosity CoreG show
a remarkable similarity to those of classical LLRG, in particular,
they share the presence of a flat core in their host's brightness profiles,
they have the same distribution in black hole
masses, as well as analogous properties concerning their optical
emission lines and radio-morphology. 
These results indicate that core galaxies and LLRG
can be considered, from these different point of view, as being drawn
from the same population of early-type galaxies. They can only be
separated on the basis of their different level of nuclear activity,
with the LLRG forming the tip of the iceberg of (relatively) 
high luminosity objects. Furthermore, the emission processes
associated to
their activity scale almost linearly over 6 orders of magnitude in
all bands for which data are available.

\begin{figure}
\centerline{
\psfig{figure=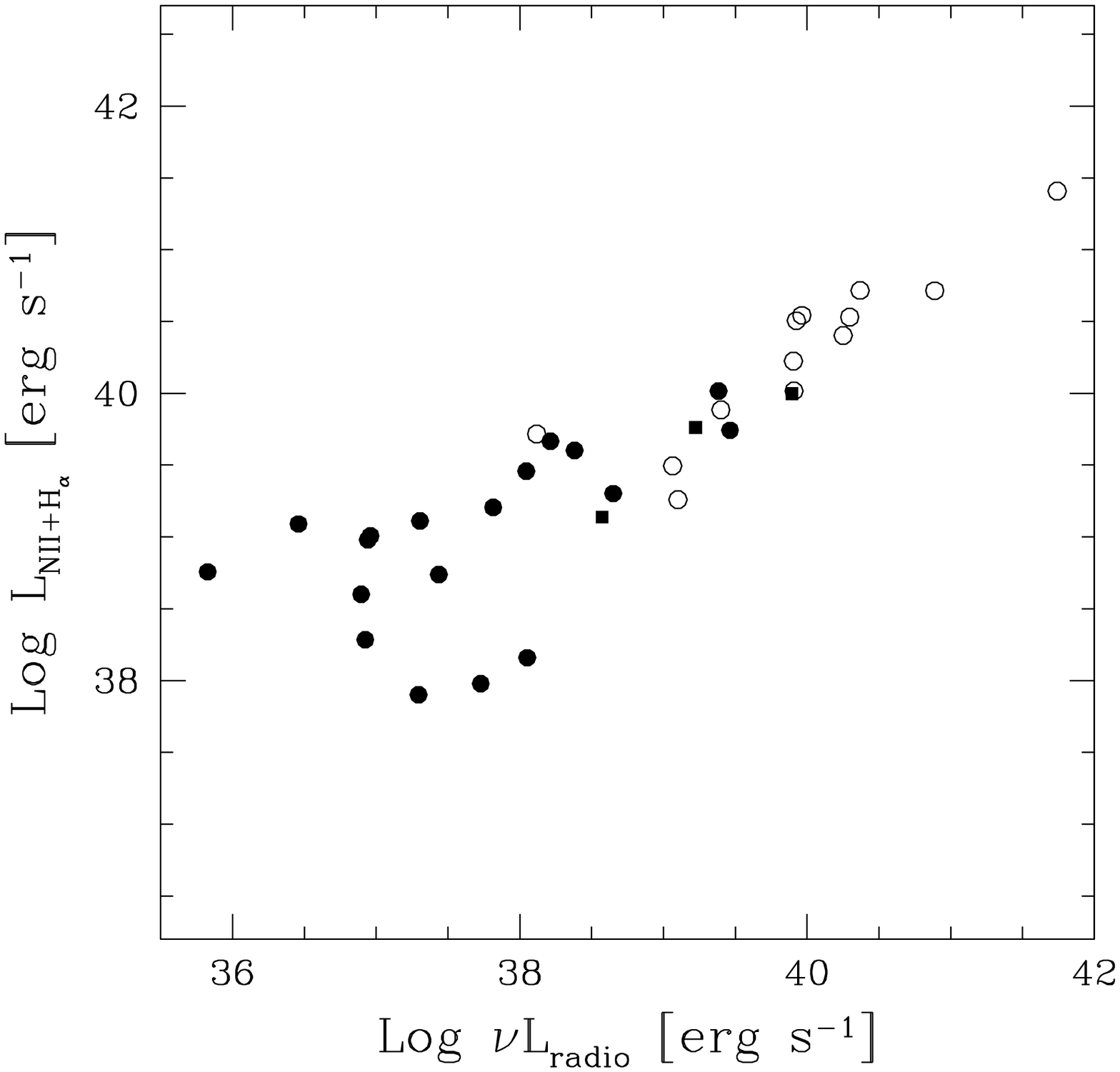,width=1.00\linewidth}
}
\caption{Emission line vs. radio core luminosity for CoreG galaxies (filled circles)
and for the LLRG 3C/FR~I sample (empty circles), from Capetti et al. (2005).}
\label{line}
\end{figure}

\section{Black hole mass and radio luminosity.}

The issue of the relationship between
the black hole mass and the radio-luminosity has been discussed by
several authors, taking advantage of the recent possibility to measure
(or at least estimate) $M_{BH}$.
\citet{franceschini98} pioneered this field showing, 
from a compilation of objects with
available black hole estimates, that the radio-luminosity tightly
correlates with the black hole mass, with a logarithmic index of
$\sim$ 2.5. This result was subsequently challenged, by
e.g. \citet{ho02}.
We here re-explore this issue limiting ourselves to the sample
of core early-type galaxies; while this substantially restricts
the accessible range in $M_{BH}$ and it applies only to radio-loud
nuclei, it has the
substantial advantage of performing the analysis on a complete sample
with well defined selection criteria and
covering a large range of radio-luminosity.

\begin{figure}
\centerline{
\psfig{figure=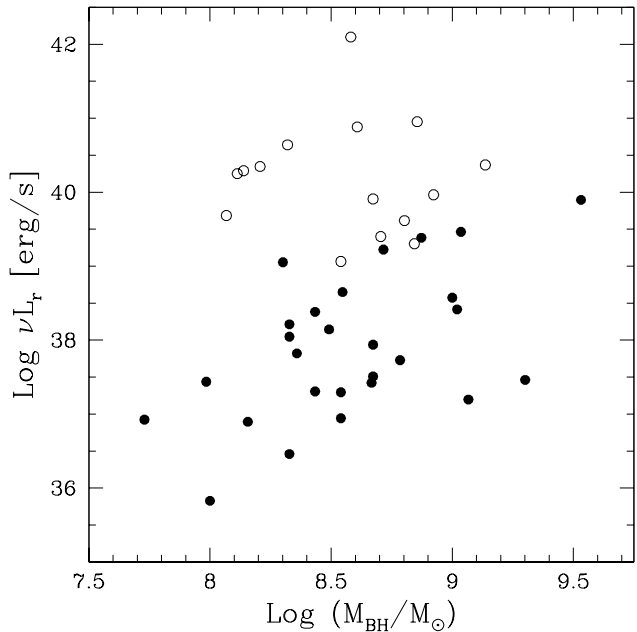,width=1.00\linewidth}
}
\caption{Nuclear radio-luminosity vs. black hole mass M$_{BH}$ for CoreG galaxies (filled circles)
and for the LLRG 3C/FR~I sample (empty circles).}
\label{mbh}
\end{figure}

The comparison of  the radio-core luminosity with the  black hole mass
is presented in  Fig.  \ref{mbh}. Apparently,
a dependence of $L_r$ on  $M_{BH}$ is present,
although   with  a  substantial   scatter. However, 
the radio flux limit of the samples exclude objects with lower radio
luminosity, potentially populating the lower part of the 
$L_R$ vs. $M_{BH}$ plane.
Furthermore, the
inclusion of LLRG (which, as discussed above, represent the high
activity end of the early-type population) radically changes the picture, as they populate the
whole upper portion of this plane. This indicates that  a very large
range (at least 4 orders  of magnitude) of radio-power can correspond to
a given $M_{BH}$ 
\footnote{With respect to previous studies we report the
nuclear   radio   emission   only,   instead  of   the   total   radio
luminosity.  However, since  the fraction  of extended  emission grows
with radio luminosity, using the  total power would just move the LLRG
upward, further increasing  the spread.}. 
This is a clear
indication that, not unexpectedly, 
parameters other than the black hole mass play a
fundamental role in determining the radio luminosity of a galaxy.

More notable is the lack of sources with 
$M_{BH} < 10^8 M_{\sun}$ 
(with only one exception). 
The  effects  produced  by  our
selection  criteria must be  considered before  any conclusion  can be
drawn.  In particular the correlation between the black hole mass and
the  spheroidal   galactic  component,  combined   with  the  limiting
magnitude,  translates into  a threshold  in the  accessible  range of
black  hole masses.  Using the limit in apparent  magnitude of our
sample ($m_B <  14$), an
average color  of B-K =  4.25 \citep{mannucci01}
and the best fit to the relationship between
$M_{BH}$ and $M_K$  from \citet{marconi03} we  obtain that at distances
larger  than  20 Mpc  (corresponding  to 7/8 of  the  volume covered
in the Wrobel's sample) we  do include galaxies with expected
black hole masses $M_{BH} < 10^8 M_{\sun}$. This represents
a severe bias against the inclusion of galaxies with low values
of $M_{BH}$, regardless of their radio emission.
The lack of low black hole mass LLRG seems to favour  
the reality of this effect, as they
are  not directly selected imposing an optical threshold; 
however, the already discussed statistical trend linking
radio and optical luminosity might represent a more subtle
bias leading to the same effect.
The existence of a minimum black hole mass
to produce a radio-loud nucleus must be properly
tested extending the analysis to a  sample  of less luminous
galaxies, likely to harbour less massive black holes.

\section{Constraints on the radiative manifestation of the accretion process}
\label{adaf}

\begin{figure*}
\centerline{
\psfig{figure=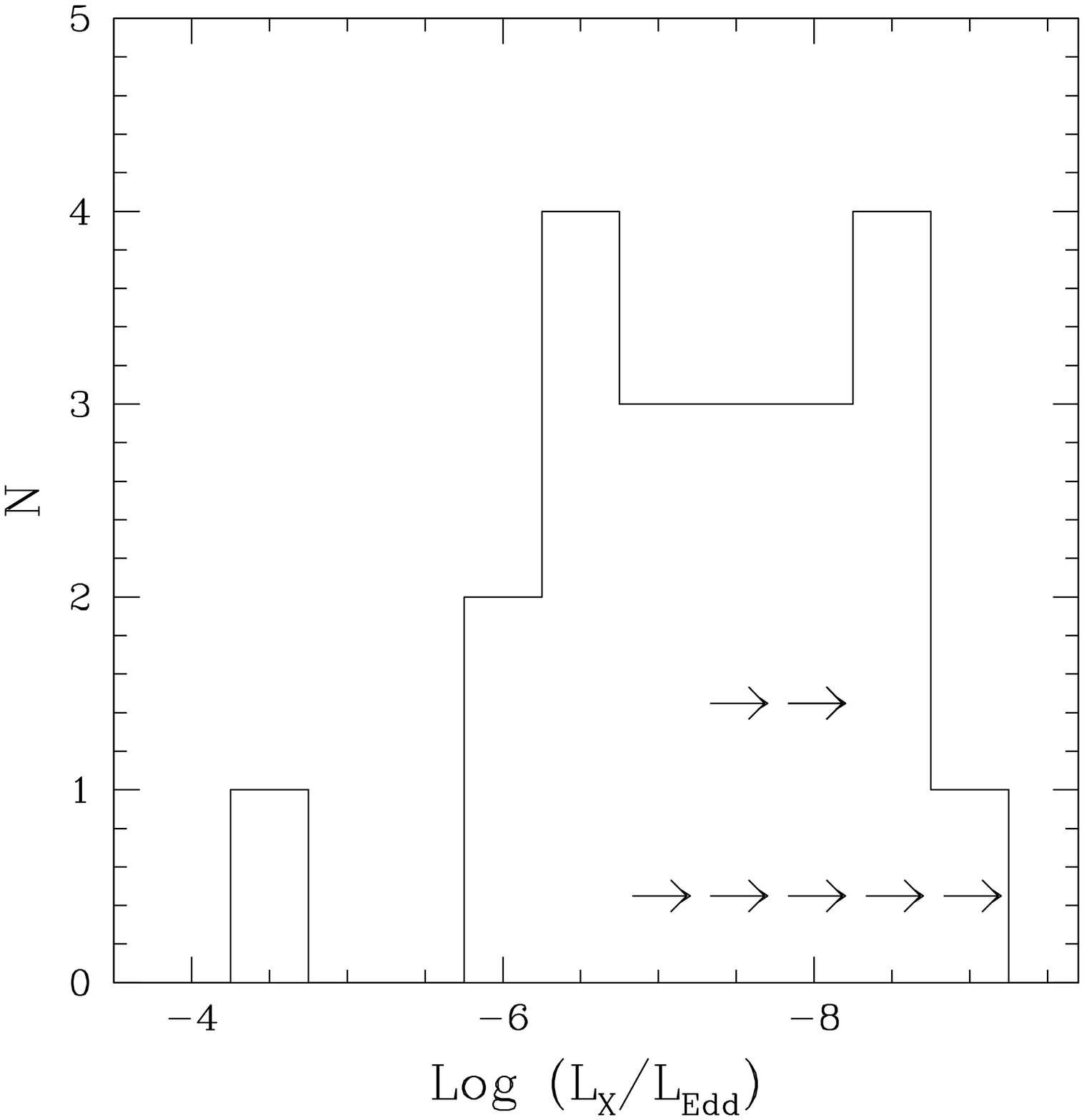,width=0.50\linewidth}
\psfig{figure=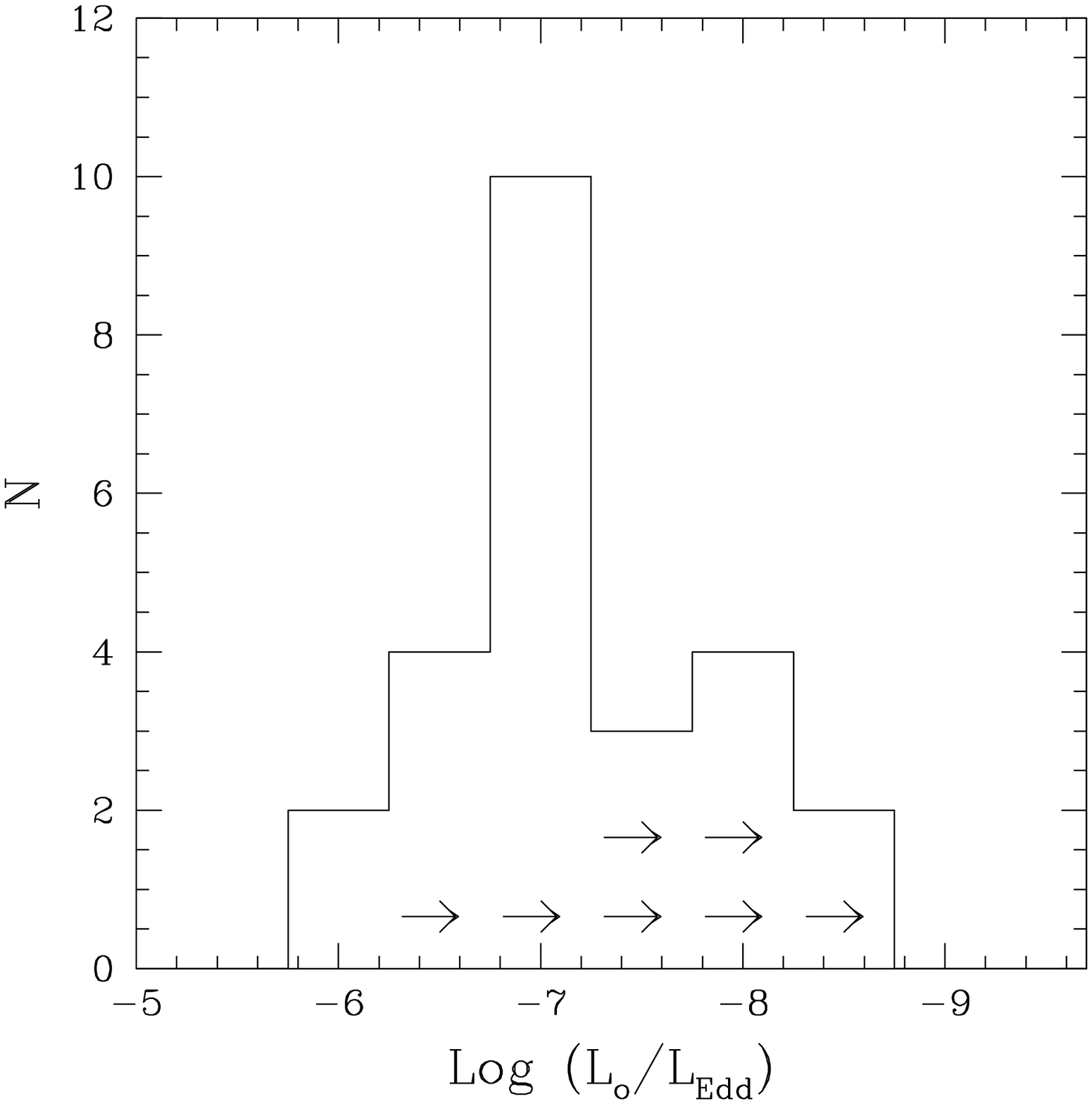,width=0.50\linewidth}
}
\caption{\label{eddihis} Distributions of the nuclear luminosities 
measured as fraction
of the Eddington luminosity in the X-ray (left) and optical (right)
bands. }
\end{figure*}

Taking advantage of the estimates of black hole mass we 
can convert the measurements of
the nuclear luminosities to units of the Eddington luminosity. 
All CoreG nuclei are associated with a low fraction of $L_{\rm {Edd}}$, 
being confined to the range $L/L_{\rm {Edd}} \sim 10^{-6} - 10^{-9}$
in both the X-ray and optical bands (with only one X-ray exception), 
see Fig. \ref{eddihis}. 
Furthermore, as discussed in Sect. \ref{nuclei},
the tight correlations between radio, optical and X-ray nuclear 
luminosities extending across LLRG and CoreG strongly argue in favour of a jet
origin for the nuclear emission also in the core galaxies.
If this is indeed the case, the observed nuclear emission 
does not originate in the accretion process and the
values reported above should be considered
as upper limits.

Our results add to the already vast literature reporting 
emission corresponding to a very low Eddington fraction associated with 
accretion onto supermassive black holes. 
These results prompted the idea that in these objects accretion occurs
not only at a low rate but also at a low radiative efficiency,
such as in the Advection Dominated Accretion Flows 
\citep[ADAF,][]{narayan95} in which most of the
gravitational energy of the accreting gas is advected into the
black hole before it can be dissipated radiatively, thus reducing the
efficiency of the process with respect to the standard models of
geometrically thin, optically thick, accretion disks.
The ADAF models have been rather successful in modeling the observed
nuclear spectrum in several galaxies, such as e.g. the
Galactic Center and NGC 4258 \citep{narayan95,lasota96}. 
Conversely, ADAF models substantially 
over-predict the observed emission in the nuclei of nearby bright elliptical
galaxies \citep{dimatteo00,loewenstein01}.

This suggested the possibility that a substantial fraction
of the mass included within the Bondi's accretion radius
\citep{bondi52} might not
actually reach the central object,
thus further reducing the radiative emission from the accretion
process with respect to the ADAF models. 
This may be the case in the presence of an outflow
\citep[Advection Dominated Inflow/Outflow Solutions, or ADIOS,][]{blandford99}
or strong convection \citep[Convection Dominated Accretion Flows,
  or CDAF,][]{quataert00} in which most gas circulates in 
convection eddies rather than accreting onto the black hole. 

Unfortunately, in the case of the galaxies under investigation,
the comparison of the theoretical predictions 
with the observations so as to get constraints
on the properties of the accretion process is quite difficult. 
This is due to the presence of different competing models, all
of these with several free parameters, and to the observational data, in particular to
the scarce multiwavelength coverage of the nuclear emission
measurements which
prevents us from deriving a detailed Spectral Energy Distribution of these objects.   
As discussed above, this is more complicated for our radio-loud
nuclei in which  the emission is most likely dominated by the
non-thermal radiation from their jets.

Nonetheless, \citet{pellegrini05} recently studied in detail a sample
of nearby galaxies for which the Chandra observations provide an
estimate of the temperature and density of the gas in the nuclear
regions, thus enabling one to derive the expected Bondi 
accretion rate, $\dot{M}_B$. It is interesting to note that the estimates 
of $\dot{M}_B$ for three sources common to both samples
with the lowest X-ray luminosity (namely NGC~1399, UGC~7629 (AKA
NGC~4472) and UGC~7898 (AKA NGC~4649)) are relatively large, 
$\dot{M}_B/\dot{M}_{\rm Edd} = 10^{-2} - 10^{-4}$,
while their X-ray luminosities are $L_X/L_{Edd} = 10^{-8} - 10^{-10}$
(see her Fig. 3). These luminosities are between
3 and 5 orders of magnitude lower than expected from an
ADAF model, and they should be considered
only as upper limits. 
These results argue in favour of an effective accretion rate
substantially smaller than expected in the case
of spherical accretion, suggesting that 
an important role is played by mass loss due to an outflow or 
by convection.

\section{CoreG and the BL~Lacs/LLRG unifying model}
\label{bllac}

\begin{figure}
\centerline{
\psfig{figure=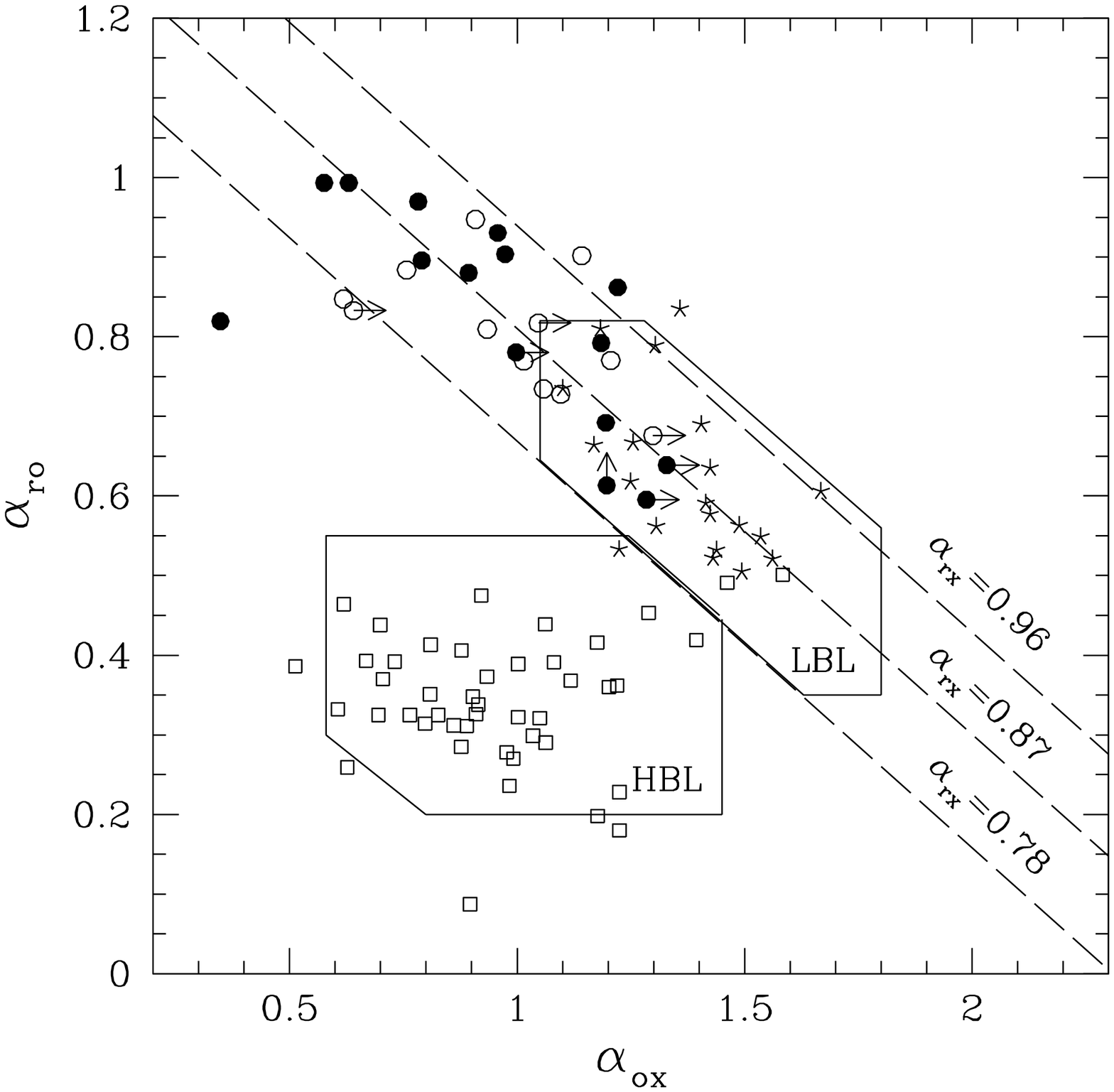,width=1.00\linewidth}
}
\caption{Broad band spectral indices, calculated 
between 5 GHz, 5500 \AA\ and 1 keV, for core galaxies (filled
circles), low luminosity 3C/FR~I radio-galaxies (empty circles), 
Low energy peaked BL Lacs (stars) and High energy peaked BL Lacs
(squares).
Solid lines mark the regions within 2
$\sigma$ from the mean $\alpha_{ro}$ and $\alpha_{ox}$ for BL Lacs 
drawn from the DXRB and RGB surveys.
The dashed lines
represent constant values for the third index, $\alpha_{rx}$.}
\label{roox}
\end{figure}

\begin{figure*}
\centerline{
\psfig{figure=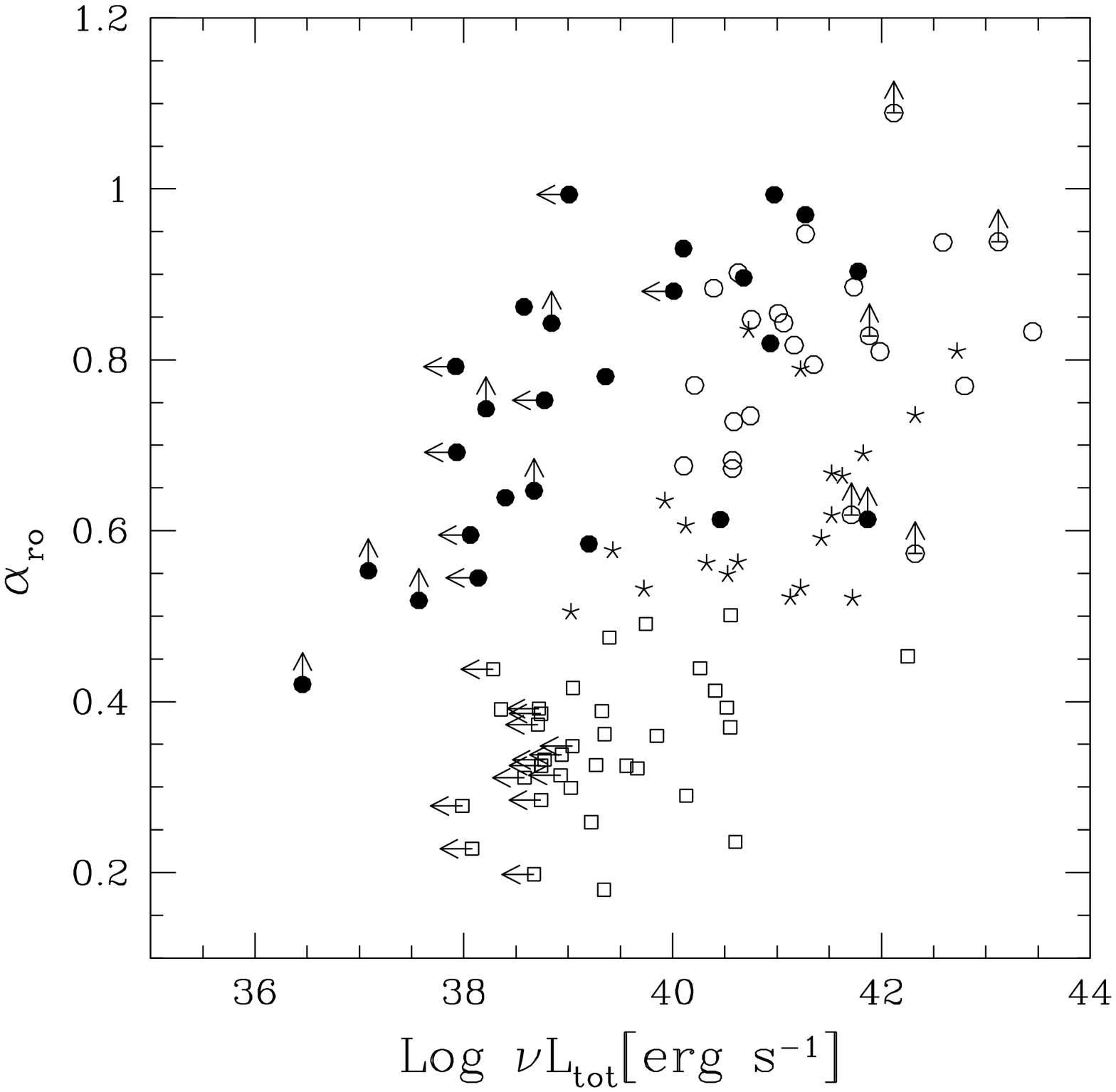,width=0.50\linewidth}
\psfig{figure=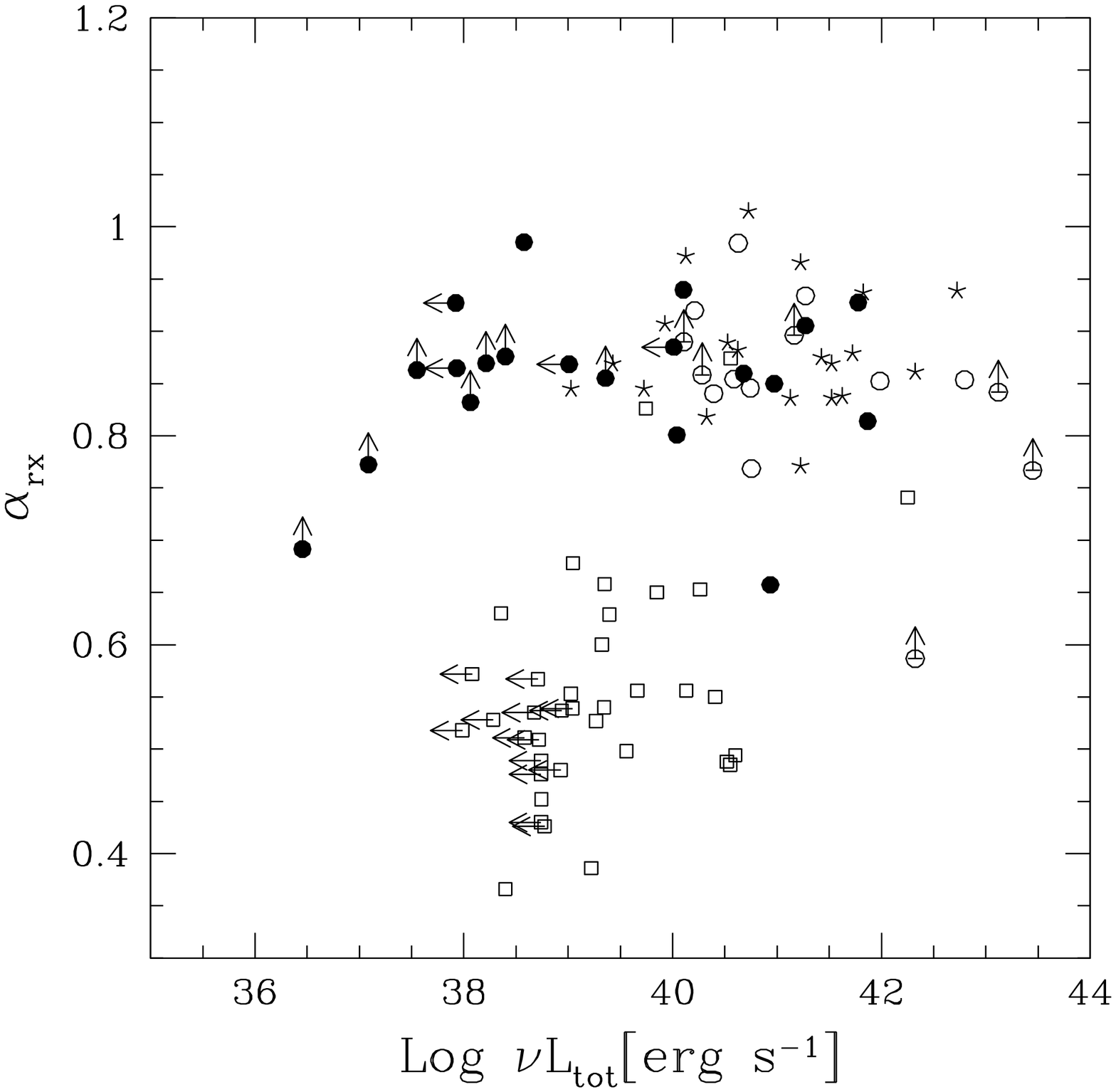,width=0.50\linewidth}
}
\caption{\label{spix} 
Broad band spectral indices vs. extended radio luminosity 
for core galaxies (filled
circles), LLRG (empty circles), 
LBL (stars) and HBL (squares). The CoreG luminosity has been extrapolated
to 178 MHz adopting a spectral index of 0.7.}
\end{figure*}

In Section \ref{cg-fri} we presented evidence that 
``core'' galaxies and LLRG are drawn from the same
population of early-type galaxies. They can only be
separated on the basis of their different level of nuclear activity,
with CoreG representing the low luminosity extension of LLRG. 
The CoreG nuclei appear to be the scaled down versions of those of LLRG
when their multiwavelength nuclear properties are considered. 
Thus here we are sampling a new 
regime for radio-galaxies in terms of nuclear power and
it is important to explore the implications of this result 
for the model unifying BL Lac objects and radiogalaxies.

Unification models ascribe the differences between the observed
properties of different classes of AGN to the anisotropy of the
nuclear radiation 
\citep[see e.g.][ for reviews]{antonucci93,urry95}.
In particular, for low
luminosity radio-loud objects, it is believed 
that BL Lac objects are the pole-on counterparts of radio-galaxies,
i.e. their emission is dominated by the radiation 
from the inner regions of a relativistic jet seen 
at a small angle from its axis which is thus strongly
amplified by relativistic Doppler beaming. 
In FR~I, whose jets are
observed at larger angles with respect to the line of sight,
the nuclear component is strongly de-amplified.  
Contrary to other classes of AGN there is growing evidence that
obscuration does not play a significant role in
these objects \citep{henkel98,chiaberge:ccc,donato04,balmaverde05}.

\citet{balmaverde05} found that there is a close similarity of the
broad band spectral indices between LLRG and the 
sub-class of the BL~Lacs, the Low energy peaked BL~Lac 
\citep[LBL, ][]{padovani95}, in agreement with the
unified model\footnote{The small offsets between the two classes
can be quantitatively 
accounted for by the effects of beaming since
Doppler beaming not only affects the angular pattern
of the jet emission, but it also causes a shift in frequency of the
spectral energy distribution \citep[see ][]{marco3,trussoni03}}.
We performed the same comparison 
\footnote{We used the standard definition of spectral indices,
measured between 5 GHz, 5500 \AA\ and 1 keV. Optical fluxes have been
converted from 8140 \AA\ to 5500 \AA\ using a local slope of $\alpha =
1$; 1 keV fluxes are
directly derived from the spectral fit.} including CoreG,
see Fig. \ref{roox}. 
We considered the radio selected BL Lacs sample
derived from the 1Jy catalog \citep{stickel91}
and the BL Lac sample selected from the
{\it Einstein} Slew survey \citep{elvis92,perlman96}. 
We used the classification into 
High and Low energy peaked BL Lacs (HBL and LBL respectively), as well
as their multiwavelength data 
given by \citet{fossati98}. We also report the regions 
(solid lines) of the plane within 2
$\sigma$ from the mean $\alpha_{ro}$ and $\alpha_{ox}$ for the BL Lacs 
drawn from the Deep X-Ray Radio Blazar Survey (DXRBS) 
and the ROSAT All-Sky Survey-Green Bank Survey (RGB) \citep{padovani03}.

Core galaxies are found to be located in  the same region covered by  LLRG. 
This   is  not  surprising  since  they extend
the behaviour of LLRG in  the radio/optical and radio/X-ray planes,
following Log-Log linear correlations whose slope is close to
unity, implying only a small dependence of spectral indices on
luminosity. More importantly, they populate the same area
in which LBL are found.

We also compared 
the spectral indices of the different groups taking into
account the extended radio-luminosity L$_{ext}$
(see Fig. \ref{spix}) which does not depend on orientation. This enables us 
to properly relate objects from the same region of the luminosity
function of the parent population. 
Indeed, the strongest evidence in favour of 
the FR~I/BL~Lac unifying model comes from the similarity in
the power and morphology of the
extended radio emission of BL Lacs and FR~I 
\citep[see e.g.][]{antonucci85,kollgaard92,murphy93}.

The CoreG reach radio-luminosities $\sim$ 100 smaller 
than in LLRG and the 1 Jy LBL.
In addition, in 13 CoreG the available radio-maps do not allow us to separate
core and extended radio-emission and $L_{ext}$ must be considered as
an upper limit.
This suggests that the CoreG represent the counterparts of 
the large low luminosity population of BL~Lac of LBL type which is
now emerging from the low radio flux limit surveys such as the DXRBS 
\citep{landt01}. Clearly, this still requires measurements of the
extended radio-luminosity of these low power BL~Lac.
A ramification of this possible extension of the unified model
toward lower luminosities 
would be the presence of relativistic jets also in our 
sample of quasi-quiescent early-type galaxies, as this is a prerequisite 
to produce a substantial dependence of the luminosity on the viewing angle.

We did not find any CoreG with spectral properties similar
to those of the High energy peaked BL~Lac (HBL), even though HBL have
extended radio-emission values of $L_{ext}$ similar to CoreG.
The spectral indices of CoreG imply 
a difference  in both the radio-to-optical and
radio-to-X-ray flux ratios of an average factor 
of  $\sim$100 with respect to HBL.
The same result applies to LLRG, as all have a LBL-type SED, with the
only exception of 3C~264 \citep{fr1sed}.
Our optical selection criteria did not exclude the  parent population of  HBL 
since their host galaxies are early-type  sufficiently luminous 
\citep[$M_R < -22.5$,][]{scarpa00} to be included  in our
sample. Most likely, the dearth of HBL-like CoreG is induced by the radio
threshold. Purely radio selected
samples of BL~Lacs are known to strongly favour the inclusion of LBL; 
e.g. in the 1 Jy sample
there are only 2 HBL out of 34 objects \citep{giommi94}.

\section{Summary and conclusions}

The aim of this series of papers is to explore
the classical issue of the connection between host galaxies and AGN,
in the new light shed by the recent developments 
in our understanding of the nuclear regions of
nearby galaxies. 

We thus selected a samples of nearby early-type 
galaxies comprising 332 objects. We performed an initial
selection of AGN candidates requiring a radio detection 
above $\sim$1 mJy leading to a sub-sample of 112 sources.
Archival HST images enabled us to classify 51 of them
into core and power-law galaxies on the basis
of their nuclear brightness profile.
We here focused on the 29 core galaxies.

We used HST and Chandra archival data to isolate their nuclear
emission in the optical and X-ray bands, 
thus enabling us (once
combined with the radio data) to study the
multiwavelength behaviour of their nuclei.
The detection rate of nuclear sources is 18/29 in
the optical (62 \%, increasing to 72\% if the sources affected by large
scale dust are not considered) and 14 in the X-ray, out 
of the 21 objects with available Chandra data (67 \%).
Our selection criteria required a radio detection 
in order to select AGN candidates;
26 CoreG are confirmed as genuine active galaxies 
based on the presence of i) an optical (or X-ray) core, ii) a AGN-like
optical spectrum, or iii) radio-jets, with only 3 exceptions,
namely UGC~968, UGC~7898 and NGC~3268.

The most important result of this analysis is that
``core'' galaxies invariably host a radio-loud nucleus.
The radio-loudness parameter $R$ for the nuclei
in these sources is on average Log R $\sim$ 3.6, a factor 
of 400 above the classical threshold between radio-loud and
radio-quiet nuclei. The X-ray data provide a completely independent
view of their multiwavelength behaviour leading to the same result,
i.e. a large X-ray deficit, at the same radio luminosity, 
when compared to radio-quiet nuclei.

Considering the multiwavelength nuclear diagnostic planes, 
we found that optical and X-ray nuclear luminosities are 
correlated with the
radio-core power, reminiscent of the behaviour
of low luminosity radio-galaxies. The inclusion of CoreG
indeed extends the correlations reported for LLRG toward much lower
power, by a factor of $\sim 1000$.

The available radio maps show that in 17 CoreG
the extended radio morphology is clearly indicative of a collimated outflow,
in the form of either double-lobed structures or jets, 
although in several cases this can only be seen in high
resolution VLBI images. This finding, combined with the 
analogy of the nuclear properties, leads us to the conclusion that
miniature radio-galaxies are associated with all core galaxies
of our sample. 

The similarity between CoreG and classical low luminosity
radio-galaxies extends to other properties. Recent results show that
LLRG are always hosted by early-type galaxies with a shallow cusp in their
nuclear profile, and this is the case, by definition, for our CoreG.
While the distributions of black hole masses, $M_{BH}$,
of the two classes are indistinguishable,
hosts of 3C/FR~I radio-sources are on average slightly more 
luminous than CoreG but there is
a substantial overlap between the two groups.
CoreG and LLRG also share similar properties from the point of view of
their emission lines, as all sources with available data conform
to the definition of a LINER on the basis of the optical line ratios
and they follow a common dependence of line luminosity
with radio core power.
CoreG and LLRG thus appear to be drawn
from the same population of early-type ``core'' galaxies. They host
active nuclei with the same multiwavelength characteristics
despite covering a range of 6 orders of magnitude in
luminosity. Thus LLRG represent the tip of the iceberg of (relatively) 
high luminosity objects.

It is unclear what mechanism is driving the level of nuclear
activity. As noted above, there is a marginal difference (less than 1
mag) in the host
galaxies of CoreG and LLRG; this reflects the well known (but as yet
unexplained) trend for which 
a brighter galaxy has a higher probability of being a stronger
radio emitter. As described in \citetalias{capetti05},
this effect is present also within our sample of CoreG
but it cannot
be simply described as a correlation between  L$_r$ and  M$_K$.

We explored if there is a relationship between the black
hole mass and the radio-luminosity.
Again, a  very large  range of
radio-power corresponds  to a  given $M_{BH}$. We do not  find any
relationship   between  radio-power  and   black  hole   mass, clearly
indicating  that parameters  other than  the  black hole  mass play  a
fundamental role in determining the  radio luminosity of a galaxy.  No
sources with $M_{BH} < 10^8  M_{\sun}$ are found.  However, this might
be due to a  bias induced by the  sample's selection criteria. 
The limit in optical  magnitude translates  into a
threshold of  accessible black  hole masses.  
Only by extending  this study to a sample of less
luminous  galaxies (harbouring,  on average,  smaller black  holes) 
will it be possible  to test the reality of a minimum  black hole mass to
produce a radio-loud nucleus.

Our data can also be used to set constraints 
on the radiative manifestation of the accretion process.
The nuclear luminosities of CoreG correspond, in units of the
Eddington luminosity, to the range $L/L_{\rm {Edd}} \sim 10^{-6} -
10^{-9}$ in both the optical and X-ray bands. 
In analogy with the scenario
proposed for LLRG, the available data support a common
jet origin for the nuclear emission in these observing bands also for CoreG.
Thus, the above values should be considered as upper limits to
the radiative manifestation of the accretion process, suggesting
that accretion occurs both at a low accretion level and at a low efficiency.
It is difficult to derive from these results clear 
constraints on the properties of the accretion flow. In part this is due to
the limited information on the Spectral Energy Distribution of the CoreG
nuclei and by the fact that in these
radio-loud nuclei the observed emission is most likely dominated by 
radiation from their jets rather than from the accretion.
This is further complicated by 
the presence of several competing accretion models
whose predictions of the emitted spectra depend on parameters that
are not well constrained by the observations.
Nonetheless, in the galaxies with the least luminous nuclei, 
the estimates of the accretion rate  from the literature
(derived for the case of spherical accretion), combined with
the very low level of X-ray emission, suggest that 
an important role is played by outflows (or 
by convection) in order to substantially suppress
the amount of gas actually reaching the central object.

As reported above, the CoreG can be effectively considered
as miniature radio-galaxies, in terms of nuclear luminosity,
thus we are sampling a new 
region in terms of luminosity for radio-loud AGN.
It is interesting to explore the implications of this result 
also for the model unifying BL Lac objects and radio-galaxies.
The broad band spectral indices of CoreG present a very close similarity
to those of Low Energy Peaked BL Lac, suggesting the extension
of the unified models to these lower luminosities. The CoreG might 
represent the mis-aligned counterpart of the large population of low
luminosity BL Lac emerging from the recent
surveys at low radio flux limits. Clearly, a more detailed comparison,
taking into account e.g. the (as yet not available) information on the
extended radio power and morphology, is needed before this result
can be confirmed. An important ramification 
of this possible extension of the unifying model
toward lower luminosities would be the presence 
of relativistic jets, 
the essential ingredient of this model, 
also in our quasi-quiescent early-type galaxies.

In the third paper of the series we will explore the properties
of the AGN hosted by galaxies with a power-law brightness profile.

\begin{acknowledgements}
This work was partly supported by the Italian MIUR under 
grant Cofin 2003/2003027534\_002.
This research has made use of the NASA/IPAC Extragalactic Database (NED)
(which is operated by the Jet Propulsion Laboratory, California Institute of
Technology, under contract with the National Aeronautics and Space
Administration), of the NASA/ IPAC Infrared Science Archive
(which is operated by the Jet Propulsion Laboratory, California
Institute of Technology, under contract with the National Aeronautics
and Space Administration) and of the LEDA database.
\end{acknowledgements}

\appendix

\section{Notes on the X-ray observations of the individual sources}
\label{notes}

We list references and provide comments for the Chandra data
and X-ray nuclear measurements found in the literature.
We also give images and spectra for the two newly detected
X-ray nuclei, as well as a summary of the results of the data analysis
which can be found in Table \ref{newchandra}.

\begin{figure}
\centerline{
\psfig{figure=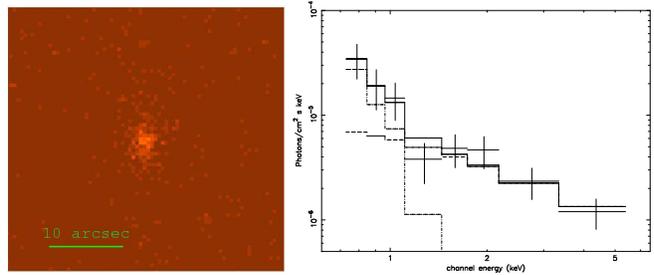,width=0.42\linewidth}
\psfig{figure=fa1b.ps,width=0.54\linewidth,angle=270}}
\caption{\label{fit1}
Chandra  image and  spectrum  for NGC~3557.
The fit and  the contributions of  the two
components (thermal and power-law) are also plotted.}
\end{figure}

\begin{figure}
\centerline{
\psfig{figure=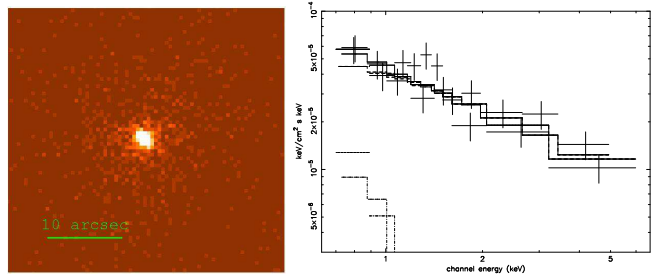,width=0.42\linewidth}
\psfig{figure=fa2b.ps,width=0.54\linewidth,angle=270}}
\caption{\label{fit2} 
Chandra  image and  spectrum  for NGC~5419.
The fit and  the contributions of  the two
components (thermal and power-law) are also plotted.
Two datasets were fit simultaneously.}
\end{figure}

\begin{table*}
\caption{
\label{newchandra} Summary of the Chandra data analysis for the two newly detected
X-ray nuclei}
\begin{tabular}{l | l l c c | c c c c c}
\cline{1-10} 
& \multicolumn{4}{|c|} {Observation information} & \multicolumn{5}{c}{Fit results} \\
Name & Obs Id & date & Inst & Exp time & N$_{H,gal}$ & $\Gamma$ & KT & F$_{x,nuc}$(1 keV) & $\chi^2$/d.o.f or PHA bins \\
\hline
NGC~3557  &   3217    & 2002-11-28    & ACIS-I & 40    & 7.4E20 & 1.1$_{-0.8}^{+0.7}$ & 0.3$_{-0.2}^{+0.5}$ & 1.1$_{-0.7}^{+1.0}E-14$ & 8 \\
\hline
NGC~5419  & 4999      & 2004-06-18    & ACIS-I & 15    & 5.46E20& 1.9$_{-0.3}^{+0.6}$ & 0.2$_{-0.1}^{+1.4}$ & 7.5$_{-1.3}^{+6.2}E-14$ & 17.86/14 \\
          & 5000      & 2004-06-19    & ACIS-I & 15    &                              &                     &                         &         \\
\hline
\end{tabular}
\end{table*}

\noindent
{\bf UGC~7360:} this object is part of the 3C/FR~I sample
of low luminosity radio-galaxies (3C~270). The Chandra data
are presented in \citet{balmaverde05}.

\noindent
{\bf UGC~7386:} 
A total of 310 nuclear counts in  0.2-8 keV band 
were  extracted from a  2\arcsec\ diameter
circle   without  background  subtraction   and  converted   to  X-ray
luminosity (2-10  keV) assuming  an intrinsic power-law  spectrum with
photon  index  $\Gamma=1.8$  and  a column  density  $N_{H}=2\;10^{20}$
cm$^{-2}$ \citep{ho01b}.

\noindent
{\bf UGC~7494:}  
this object is part of the 3C/FR~I sample
of low luminosity radio-galaxies (3C~272.1). The Chandra data
are presented in \citet{balmaverde05}.

\noindent
{\bf UGC~7629:}
Both \citet{soldatenkov03} and \citet{maccarone03} 
detected emission from the nucleus at energies
below 2.5 keV. \citet{biller04} confirmed the presence of 
nuclear emission in the 0.3-10  keV band extracting 
64 source counts in an 1\arcsec\ circle. The spectrum was modeled 
using a power-law model  with photon index $\Gamma=1.7$
and a column density $N_H=1.7\;10^{20} $cm$^{-2}$.
These results do not contrast the upper limit given by
\citet{loewenstein01} who did not find a
nuclear component in the hard X-ray band 
from 2 to 10 keV. 

\noindent
{\bf UGC~7654:}  
this object is part of the 3C/FR~I sample
of low luminosity radio-galaxies (3C~274). The Chandra data
are presented in \citet{balmaverde05}.

\noindent
{\bf UGC~7760:}
\citet{filho04} find a bright X-ray nuclear source,spatially
coincident with the radio core position.
The spectrum of the nuclear source (1200 net counts) is
well fitted by a two component model, i.e. a power law ($\Gamma=1.51$)
plus Raymond-Smith thermal plasma (KT=0.95 keV).

\noindent
{\bf UGC~7878:}
\citet{loewenstein01} give a 3 $\sigma$
upper limit to any nuclear X-ray point source converting the nuclear counts
(16 counts in an 1\arcsec\ circle region)
to luminosity in the 2-10 keV band assuming a power-law spectrum 
with a slope $\Gamma = 1.5$. 

\noindent
{\bf UGC~7898:}
\citet{soldatenkov03} detected nuclear emission at a 3 confidence level
only in the range 0.2-0.6 keV with a luminosity of $6\;10^{37}$ erg s$^{-1}$.
Conversely \citet{randall04} did not find 
conclusive evidence for a central AGN. They give an
upper limit for the X-ray luminosity converting count rate (0.3-10
keV) 
into the un-absorbed luminosity
L$_x$(0.3-10 keV) using photon index $\Gamma=1.78$.
We adopted the latter, more conservative, estimate.

\noindent
{\bf UGC~9706:}
\citet{filho04} detected a weak hard x-ray nucleus (total counts )
and fitted the
spectrum with a power-law model ($\Gamma=2.29$). 
Their result is consistent with that of \citet{trinchieri02}.

\noindent
{\bf UGC~9723:}
\citet{terashima03} did not detect an X-ray nucleus and suggested
that this object is likely to be heavily obscured. 
They set a nuclear flux upper
limit assuming the Galactic absorption column density  
and a power-law model with
photon index $\Gamma=2$.

\noindent
{\bf NGC~1316:}
\citet{kim03} detected a low luminosity X-ray AGN;
the nuclear spectrum is well reproduced by a power-law model $\Gamma=1.76$
plus a MEKAL model. 

\noindent
{\bf NGC~1399:}
\citet{loewenstein01} 
did not find a nuclear point source and they give a 3 $\sigma$
upper limit to any nuclear X-ray point source converting nuclear counts
(28 counts in an 1\arcsec\ circle region)
to luminosity in (2-10 keV band) assuming a slope 1.5 power-law spectrum .

\noindent
{\bf NGC~4696:}
\citet{satyapal04} extracted 39
events in the 2-10 keV band with a 2\arcsec\ radius
centered on the nucleus and converted to
2-10 keV X-ray luminosity assuming an intrinsic power-law spectrum
with  $\Gamma=1.8$ and a Galactic column density.

\noindent
{\bf NGC~5128:}
In \citet{kraft00} the nuclear spectrum is modeled 
assuming a power-law spectrum with a photon index of 1.9 
and N$_H=10^{23}$ cm$^{-2}$.
\citet{evans04} compared different nuclear observations of Cen A and
confirmed that the spectrum is well fitted by a heavily
absorbed power-law model, consistent with the previous observation.
 
\noindent
{\bf IC~1459:}
\citet{fabbiano03} detected an
un-absorbed nuclear X-ray source ($\sim$ 6500 total counts), 
with a power law slope $\Gamma=1.88$.

\noindent
{\bf IC~4296:} 
\citet{pellegrini03} found point-like and hard X-ray emission,
well described by a moderately absorbed ($N_H=1.1\;10^{22} $cm$^{-2}$)
power-law with $\Gamma=1.48$.

\section{High resolution radio core measurements}
\label{radionuc}

In Sect. \ref{nuc} we presented measurements of the radio core fluxes
obtained from observations obtained at higher resolution and/or
frequency with respect to the 5\arcsec\
resolution data available for the whole sample.
Since these data are  highly inhomogeneous and  given the
general  agreement with the  5 GHz VLA measurements, we
prefer to retain the values from the 
surveys by \citet{wrobel91b}  and \citet{sadler89}, but,
nonetheless, we always checked that using these estimates for the core fluxes  
our results are unchanged. Here we give one example, exploring
the influence on the correlation between radio and optical nuclear
luminosity: the slope of the best fit is 
increased by only 0.02, thus it is essentially unaffected.

\begin{figure}
\centerline{
\psfig{figure=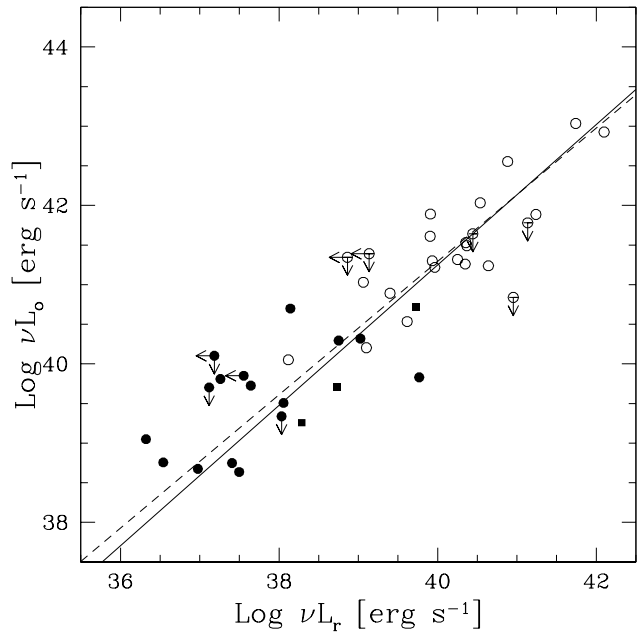,width=1.00\linewidth}
}
\caption{\label{corr_vlbi} Comparison of radio and optical 
nuclear luminosity
for the sample of core-galaxies (filled circles) using the high
resolution radio data to measure the radio cores.
Empty circles are objects from 
the 3C/FR~I sample of low luminosity radio-galaxies. 
The dashed line reproduces the best linear fit, while the solid line
shows the fits obtained with the 5 GHz VLA data, from Fig. \ref{corr1}.}
\end{figure}

\end{document}